\documentclass[aip,graphicx]{revtex4-1}

\usepackage{graphicx}%
\usepackage{multirow}%
\usepackage{amsmath,amssymb,amsfonts}%
\usepackage{amsthm}%
\usepackage{mathrsfs}%

\begin{document}

\title[]{Hydrophobically gated memristive nanopores for neuromorphic applications}

\author{Gon\c{c}alo Paulo}
\affiliation{Department of Mechanics and Aerospace Engineering, Sapienza University of Rome, Rome, 00184, Italy}
\author{Ke Sun}
\affiliation{Chemical Biology Department, Groningen Biomolecular Sciences \& Biotechnology Institute, Groningen, 9700 CC, The Netherlands}
\affiliation{Department of Laboratory Medicine, State Key Laboratory of Biotherapy and Cancer Center, Med+X Center for Manufacturing, West China Hospital, Sichuan University and Collaborative Innovation Center,  Chengdu, 610041, China}
\author{Giovanni Di Muccio}
\affiliation{Department of Mechanics and Aerospace Engineering, Sapienza University of Rome, Rome, 00184, Italy}
\author{Alberto Gubbiotti}
\affiliation{Department of Mechanics and Aerospace Engineering, Sapienza University of Rome, Rome, 00184, Italy}
\author{Blasco Morozzo della Rocca}
\affiliation{Department of Biology, Tor Vergata University of Rome,  Rome, 00133, Italy}
\author{Jia Geng}
\affiliation{Department of Laboratory Medicine, State Key Laboratory of Biotherapy and Cancer Center, Med+X Center for Manufacturing, West China Hospital, Sichuan University and Collaborative Innovation Center,  Chengdu, 610041, China}
\author{Giovanni Maglia}
\affiliation{Chemical Biology Department, Groningen Biomolecular Sciences \& Biotechnology Institute, Groningen, 9700 CC, The Netherlands}
\author{Mauro Chinappi}
\affiliation{Department of Industrial Engineering,  Tor Vergata University of Rome, Rome, 00133, Italy}
\author{Alberto Giacomello}
\email{alberto.giacomello@uniroma1.it}
\affiliation{Department of Mechanics and Aerospace Engineering, Sapienza University of Rome, Rome, 00184, Italy}

\begin{abstract}
    Signal transmission in the brain relies on voltage-gated ion channels, which exhibit the electrical behaviour of memristors, resistors with memory.
    State-of-the-art technologies currently employ semiconductor-based neuromorphic approaches, which have already demonstrated their efficacy in machine learning systems.
    However, these approaches still cannot match performance achieved by biological neurons in terms of energy efficiency and size. In this study, we utilise molecular dynamics simulations, continuum models, and electrophysiological experiments to propose and realise a bioinspired hydrophobically gated memristive nanopore. Our findings indicate that hydrophobic gating enables memory through an electrowetting mechanism, and we establish simple design rules accordingly. Through the engineering of a biological nanopore, we successfully replicate the characteristic hysteresis cycles of a memristor and construct a synaptic device capable of learning and forgetting. This advancement offers a promising pathway for the realization of nanoscale, cost- and energy-effective, and adaptable bioinspired memristors.
\end{abstract}

\maketitle

%%%%%%%%%%%%%%%%%%%%%%%%%%%%%%%%%
\section{Introduction}

With the current upsurge in the production and deployment of artificial intelligence technologies, it has become critical \cite{LeCun2019} to circumvent the bottleneck associated with processing and storing data in separate units, which is specific to the von Neumann computer architecture \cite{Sebastian2020}. Biology, which initially motivated the birth of artificial neural networks, is currently serving as a source of additional inspiration for a different paradigm in computer architectures, neuromorphic computing, which could boost the performance and sustainability of artificial intelligence \cite{Ascoli_2021,Sebastian2020,Schuman2022}.

Neuromorphic computing, as the name suggests, is shaped after the architecture of the brain, in which storage and processing of data happen in the same unit~\cite{Indiveri2015}. 
The most advanced technologies to date \cite{Strukov2008,Park2022,John2022,vandeBurgt2018,Varotto2021} implement this paradigm exploiting semiconductors; their applicability for machine learning systems has already been demonstrated \cite{Ma2017,Merolla2014}. Even though these approaches have significantly lowered the power consumption of typical neuromorphic calculations, they are still far from the performance of biological neurons~\cite{Poon2011}.

The brain indeed requires just a few watts to run and its basic operations are orchestrated by nanofluidic devices - ion channels~\cite{Alberts2002-li} -
transmembrane proteins which transmit signals in the form of ion currents. 
The non-linear behaviour that is essential for brain functions originates in the history-dependent conductance of the ion channels that are found in neurons, enabling the action potential, as first explained by Hodgkin and Huxley \cite{Hodgkin1952}. Specifically, ion channels in neurons can gate, i.e., switch on or off, depending on the transmembrane potential \cite{hille2001}.
Voltage gating typically occurs by complex action-at-a-distance mechanisms in which information is propagated from a voltage sensor domain to the ion-permeable pore, which is actuated by sterical occlusion~\cite{Bassetto2023}.

From an electrical standpoint, ion channels behave as~\emph{memristors} (memory resistors)~\cite{CHUA2012}, circuital elements whose resistance depends on the internal state of the system \cite{chua1971memristor,chua1976memristive}. 
Different architectures have been proposed to produce iontronic nanofluidic memristors~\cite{Najem2018,Robin2021,Xiong2023,robin2023}, in which ions act as charge (and information) carriers instead of electrons. Iontronics platforms have the potential of being multichannel, as their natural counterpart~\cite{Frieden2019}, with information flowing in parallel through the same circuit encoded by different ions. 

In this work we propose a hydrophobically gated memristive nanopore (HyMN), 
with an architecture inspired by biological ion channels; 
a drastic simplification is introduced in the gating mechanism, 
which relies on the formation of nanoscale bubbles to switch the ion currents thus requiring no moving parts~\cite{roth2008bubbles,giacomello2020bubble}. 
Voltage can be used to control the conductance of the nanopore imparting 
memory by electrowetting. We engineered a HyMN prototype mutating a biological nanopore, FraC. The device produces the pinched hysteresis loop in the voltage-current curve which is the signature of memristors \cite{chua1971memristor,chua1976memristive} and can behave as a synapse, learning and forgetting.
This robust and flexible design combines the advantage of being an iontronic memristor with the simplicity of a 1D system, showing promise as a basic element for innovative nanofluidic computing.

\section*{Results}

\begin{figure*}[h!]
    \centering
    \includegraphics[width=1\linewidth]{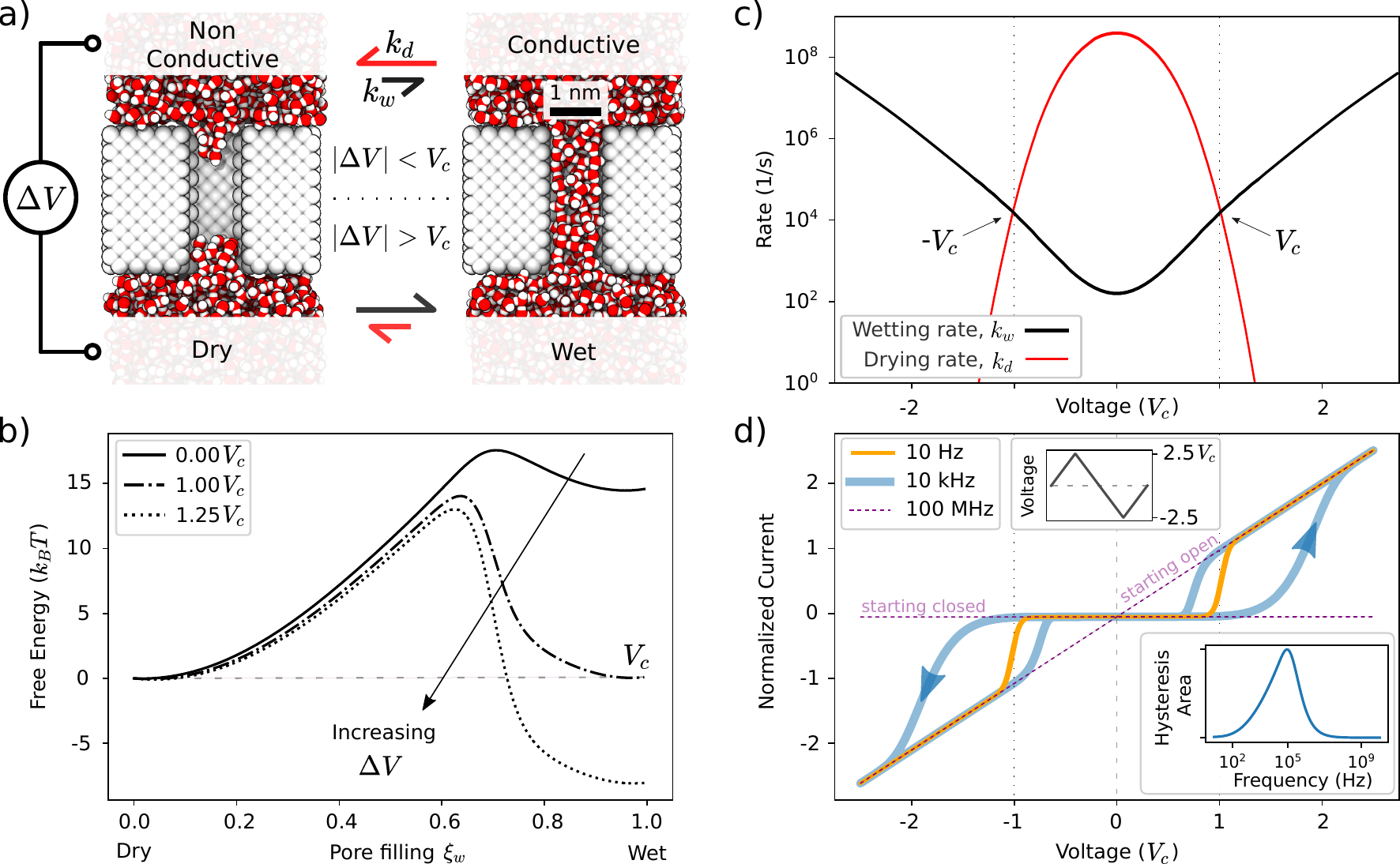}
    \caption{
    \textbf{Simple model of a memristive hydrophobic nanopore}.
    \textbf{a)} Atomistic representation of a cylindrical hydrophobic nanopore immersed in water.
    The nanopore can switch between the wet and dry state with rates $k_w$ and $k_d$ respectively. These rates depend on the applied voltage across the membrane, $\Delta V$.
    \textbf{b)} Free energy profile computed as a function of the fraction of the nanopore filled with water(water filling $\xi_w$).
    The equilibrium profile (at $\Delta V = 0$) is computed by 
    Restrained Molecular Dynamics (RMD) ~\cite{maragliano2006temperature}, 
    while the voltage dependence is estimated by using the model explained in Supp.~Note~S1.
    At $\Delta V = 0$, the most probable state 
    corresponds to the dry state. Beyond the transition voltage, 
    $\Delta V > V_c$ the wet state becomes favoured; for this specific case $V_c = 1.2V$. The error bars of the profile computed using RMD are comparable to the width of the line, 
    see Supp.~Fig.~S1.
    \textbf{c)} Variation of wetting (black) and drying rates (red) with $\Delta V$.  The drying rate, at $\Delta V = V_c$, is 5 orders of magnitude lower than its value at $\Delta V =0$, while the wetting rate is 3 orders of magnitude higher.
    \textbf{d)} 
    Current-voltage (IV) curve for an array of independent nanopores under saw-tooth voltage cycles (maximum voltage $2.5\,V_c$, top inset), at different frequencies. Three possible regimes are shown depending on the cycling frequency, going from a non-linear resistance (slow, 10 Hz) to a linear ohmic behaviour (fast, 100 MHz). At intermediate frequencies, the system shows pinched hysteresis, i.e., memory. The area of the hysteresis as a function of the cycling frequency is reported in the bottom inset, showing a maximum around 100 kHz.
}    
    \label{fig:fig1}
\end{figure*}

\subsection*{ Electrowetting of a single nanopore}
To show how hydrophobic gating can enable memristive behaviour, we consider a simple nanopore model (Fig.~\ref{fig:fig1}a), consisting of a hydrophobic cylinder with a diameter of 1 nm and a length of 2.8 nm, mimicking the sizes of biological nanopores
\cite{hille2001}.
When immersed in water, 
the nanopore lumen can be found either in the dry or in the wet states (Fig.~\ref{fig:fig1}a), due to its small size and hydrophobicity \cite{giacomello2020bubble}.
The dry state is characterised by the presence of a vapour bubble, which precludes the flow of water and ions, 
resulting in a non-conductive (gated) 
pore~\cite{smeets2008noise,roth2008bubbles,Aryal2015}.
 
The wet and dry states correspond to two different minima of the free energy, separated by a barrier. 
In the following, we will refer to the global minimum as the stable/most probable state, while  the metastable state corresponds to the local minimum.

The full (equilibrium) free energy profile, 
obtained by Restrained Molecular Dynamics (RMD) \cite{maragliano2006temperature}, 
is reported in Fig.~\ref{fig:fig1}b (solid black line), 
and in Supp.~Fig.~S1. 
For our model pore, 
the global free energy minimum corresponds to the dry (non-conductive) state;
the free energy  barrier for wetting is about $18\, k_BT$, 
while the drying one is less than $5\, k_BT$.

By applying an external voltage 
$\Delta V$ across the nanopore, 
it is possible to shift the free energy profile towards the wet state (Fig.~\ref{fig:fig1}b) thereby changing its conductance, 
for details see Supp.~Note~S1 and Supp.~Fig.~S2.
The origin of this effect is electrowetting -- 
the electric field favours the wetting of the pore by electrostricting the water meniscus~\cite{Dzubiella2005}.
The voltage at which the stable state switches from
the dry to the wet is indicated as $V_c$.
For $\Delta V > V_c$, the system is preferably in the wet, conductive state. In analogy to electronic memristors~\cite{CHUA2012}, 
the voltage-dependence of the ionic conductance of the nanopore shown in Fig.~\ref{fig:fig1}a is the crucial ingredient for developing a hydrophobically gated memristor.

In Fig.~\ref{fig:fig1}c we report the wetting and drying transition rates ($k_w$ and $k_d$, respectively) computed at different $\Delta V$, which are fundamental to assess the memory behaviour of the system; the protocol to accurately estimate these rates is discussed in Supp.~Note~S2 and Supp.~Fig.~S3. Indeed, the emergence of memory is due to the finite time that the system takes to transition from the metastable state to the stable state. 
Consider for example a pore which, 
at a moment $\tau_0$, is in the dry state:
by switching instantaneously the voltage to $\Delta V> V_c$, 
the system will remember the previous dry state for a certain time $\tau_w \simeq 1/k_w$.
In this dry state ions cannot translocate through the pore 
and the nanopore is non-conductive even if $\Delta V > V_c$.
However, if the previous condition of the system was wet, at the same voltage the nanopore would be conductive. 
In the next section, 
we will show how the dynamic modulation of the wet/dry bistability of an ensemble of HyMNs generates a pinched IV loop, the hallmark of memristors.

\subsection*{ Collective Behaviour and Pinched Hysteresis Loop}

Figure~\ref{fig:fig1}a-c shows that a single model pore can only be observed in a conductive (wet) 
or a non-conductive (dry) state.
Instead, an array (ensemble) of pores would have a distribution of wet and dry pores, whose ratio depends, inter alia on the applied voltage. The transition from single pore to the ensemble behaviour is discussed in Supp.~Fig.~S4, showing that just some tens of pores are needed to observe a continuous response as opposed to a stochastic one.

The average per-pore conductance $G=\frac{1}{N_p}\frac{I}{\Delta V}$, with  $N_p$ the number of pores, $I$ the total current, and $\Delta V$ the applied voltage at a given moment, of an ensemble of pores is given by
\begin{equation}
    G(\Delta V, t) = g_0 \, n(\Delta V, t)
    \; ,
    \label{eq:G}
\end{equation}
where $g_0$ is the single wet pore conductance and
$n$ the probability that a single pore is wet.
$n$ is history dependent and,
in the limit of an infinite number of pores, 
its evolution can be described by a master equation
\begin{equation}
    \frac{d n}{dt} = (1-n)\,k_w  -n \, k_d
    \; ,
    \label{eq:dndt}
\end{equation}
with $k_{w/d}(\Delta V)$ the voltage-dependent wetting/drying rates in Fig.~\ref{fig:fig1}c.

In Fig.~\ref{fig:fig1}d we report three current-voltage (IV) curves
obtained by the numerical integration of 
Eqs.~\ref{eq:G} and \ref{eq:dndt} 
under a saw-tooth potential at different cycling frequencies.
The picture shows that an array of HyMNs has three possible regimes:
i) at low frequencies (10 Hz, orange line), the array behaves as a non-linear resistor, because the system has enough time to visit both the wet and dry states with the equilibrium probabilities;
ii) at high frequencies (100 MHz, dashed pink), the system behaves as an ohmic resistor with finite or infinite resistance, depending on its initial wet or dry state, respectively; in this regime the voltage variation is too fast to allow to move away from the local equilibrium;
iii) at intermediate frequencies (10 kHz, blue) the system displays a pinched-loop hysteresis, i.e., memristive behaviour. This happens because the cycling frequency does not allow a complete equilibration of all the pores of the array to their stable state.
As a consequence,
the number of the wet pores at a given moment
strongly depends on the previous state, i.e., the system has memory. 
For instance, starting with all dry pores,
the total current will increase with increasing voltage,
but with some delay as compared to the equilibrium wet pore probability: 
cf. the blue and orange lines of Fig.~\ref{fig:fig1}d; 
for $\Delta V > \Delta V_c$.
%For a given voltage range ($|\Delta V| < V_\mathrm{max}$),
The inset of Fig.~\ref{fig:fig1}d shows that the memristive behaviour is observed over a rather broad range of frequencies; in this example, $10^2<f<10^7$~Hz.
A number of parameters can influence this range and the location of the maximum, see also Supp.~Fig.~S5.
Memristors can be classified in different types depending on the shape of their IV curves~\cite{robin2023,Pershin2011}, see Supp.~Fig.~S6. Our model nanopore is unipolar, which is expected based on the system symmetry.

%%%%%%%%%%%%%%%%%%%%%%%%%%%%%%%%%
\subsection*{Design criteria for HyMNs}
\label{sec:design}
The previous analysis demonstrated 
that a pinched hysteresis loop 
-- the fingerprint of memristors -- 
can be produced by an ensemble of hydrophobically gated nanopores.

Based on the physical insights into the gating mechanism, we identify four design criteria that a nanopore must satisfy to behave as an efficient HyMN:
\begin{enumerate}
\item The pore must be preferentially dry at $\Delta V = 0$;
\item The pore must undergo electrowetting before the maximum voltage $\Delta V^*$ that the system can sustain; e.g., for biological pores embedded in lipid membranes no more than 300~mV can usually be applied~\cite{yu2021stable}, while solid-state membranes can bear voltages up to some volts, depending on thickness and other parameters~\cite{yanagi2017thickness, powell2011electric};
\item The pore must dry “quickly” at $\Delta V=0$ to ensure a fast transition from the wet state to the dry state;
\item The pore must wet “quickly” at the maximum voltage $\Delta V^*$
to ensure a fast transition from the dry state to the wet state.
\end{enumerate}

\begin{figure*}[h!]
    \centering
    \includegraphics[width=1\linewidth]{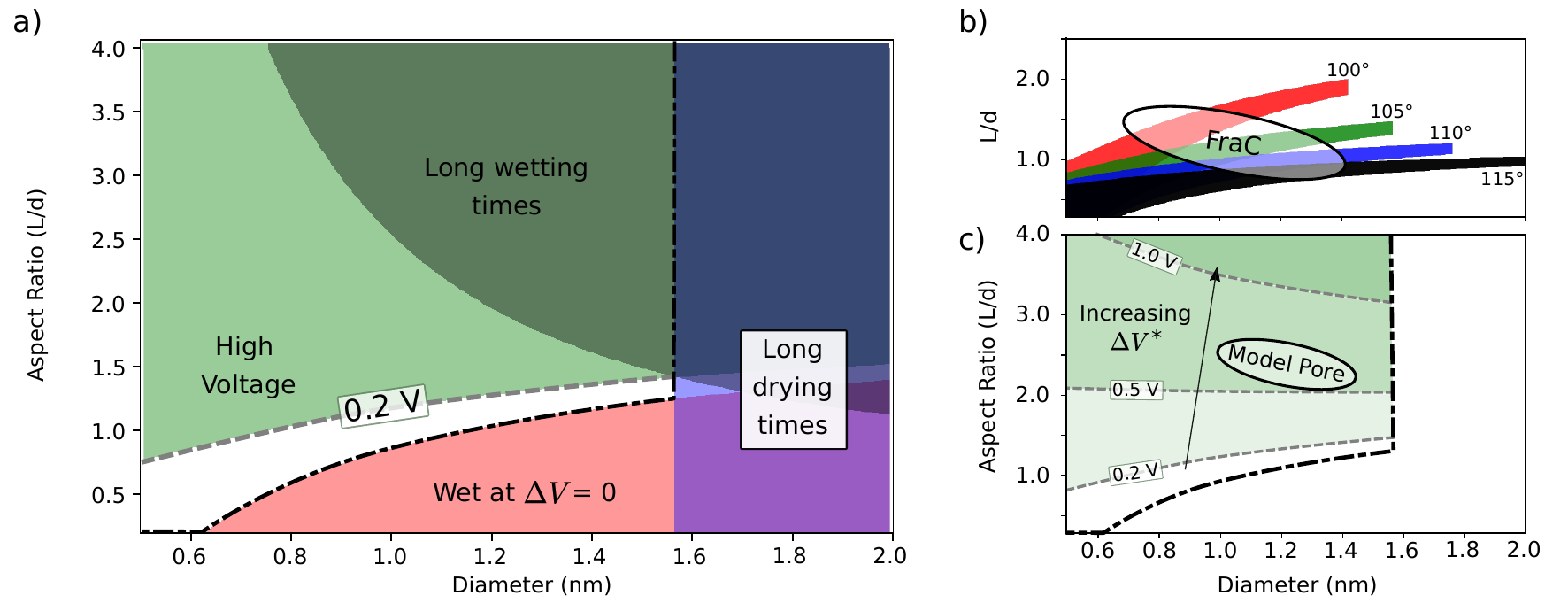}
    \caption{
    \textbf{Design criteria for HyMNs}. 
    \textbf{a)} The intersection of the 4 main design criteria exposed in the main text define a white region in the diameter vs. aspect ratio plane where hydrophobic gating is expected. Here, contact angle $104^\circ$ and maximum voltage $\Delta V^\ast = 0.2$ V are assumed. 
    The green region is forbidden as devices require voltages higher than $\Delta V^\ast$ to wet. The red region corresponds to HyMNs for which the wet state is more favourable than the dry state at 0~V. The blue region, corresponds to HyMNs that more than 10 seconds to dry when no voltage is applied, i.e. slow dryers. 
    The gray region corresponds to HyMNs 
    taking longer than 10 seconds to wet (i.e. slow wetters) at $\Delta V = 0.2$~V.
    \textbf{b)} “Allowed” regions for 4 different contact angles, from $100^\circ$ to $115^\circ$, at fixed $\Delta V^\ast = 0.2$~V. Lower contact angles give rise to slightly bigger areas, very similar aspect ratios for small diameters, but very different aspect ratios for large diameters. The ellipse shows the approximate region where a specific biological channel (FraC heptamer, see Fig.~\ref{fig:fig3}) belongs, considering the maximum and minimum radius of the hydrophobic constriction where a bubble could form. 
    \textbf{c)} Allowed regions for 
    $0.2\,\mathrm{V} < \Delta V^\ast < 1.5\,\mathrm{V}$, 
    at fixed contact angle $104^\circ$. 
    Different shades of green represent the movement of the high voltage 
    line (dashed gray) in panel a).
    Higher $\Delta V^\ast$ greatly increases the 
    allowed area (between the dashed grey and the black dash-dotted line),
    allowing for longer pore to have the memristive behaviour.
    The ellipse corresponds to the model nanopore in Fig.~\ref{fig:fig1}a.
}
    \label{fig:fig2}
\end{figure*}

The four conditions above require the fine-tuning of a non-linear combination of different physical properties of the system, like the hydrophobicity of the material, the radius and length of the nanopore, and the susceptibility of the pore to wetting by applying a voltage. To explore how different geometries and physical parameters  affect the wetting and drying dynamics,  
we constructed a macroscopic model based on classical nucleation theory to estimate the wetting and drying rates, taking into account the effect of the voltage; 
the full details of the model are described
in Supp.~Note~S3.
Within this model, 
we find that the range of parameters satisfying the previously expressed requirements is restricted to narrow (sub)nanometer-sized pores and to aspect ratios close to unity, 
see Fig.~\ref{fig:fig2}a.

The drying time depends mostly on the diameter of the pore and its hydrophobicity, while the wetting time depends also on the length of the pore. These characteristic times restrict the size of the pore to the nanoscale, as pores with larger diameters would not dry once wet and longer pores would require too high voltages to wet. 
The hydrophobicity is here quantified by the contact angle between a liquid droplet and a flat slab of the material; this is the dominant factor controlling the allowed aspect ratio to have a functioning HyMN, 
see Fig.~\ref{fig:fig2}b. 
Some biological channels fall in or near the region where hydrophobic gating is possible, and in fact some are known to do so, like CRAC~\cite{guardiani2021unveiling} and 
BK~\cite{jia2018hydrophobic} channels. For biological channel, a single contact angle does not fully characterise its hydrophobicity, but the ellipse in Fig.~\ref{fig:fig2}b shows that a mildly hydrophobic pore constriction is sufficient to achieve hydrophobic gating.
In the next section, we explore the biological FraC channel, whose approximate dimensions are represented by a white ellipse in Fig.~\ref{fig:fig2}b.
When allowing for higher maximum voltages $\Delta V^*$,
the range of aspect ratios can be significantly expanded, 
see Fig.~\ref{fig:fig2}c; 
the ellipse denotes the approximate position in parameter space of the model pore in Fig.~\ref{fig:fig1}, 
which indeed displays wetting around $\Delta V = 1.2$~V.

\subsection*{A biological HyMN: the engineered FraC nanopore}

\begin{figure*}[h!]
    \centering
    \includegraphics[width=1\linewidth]{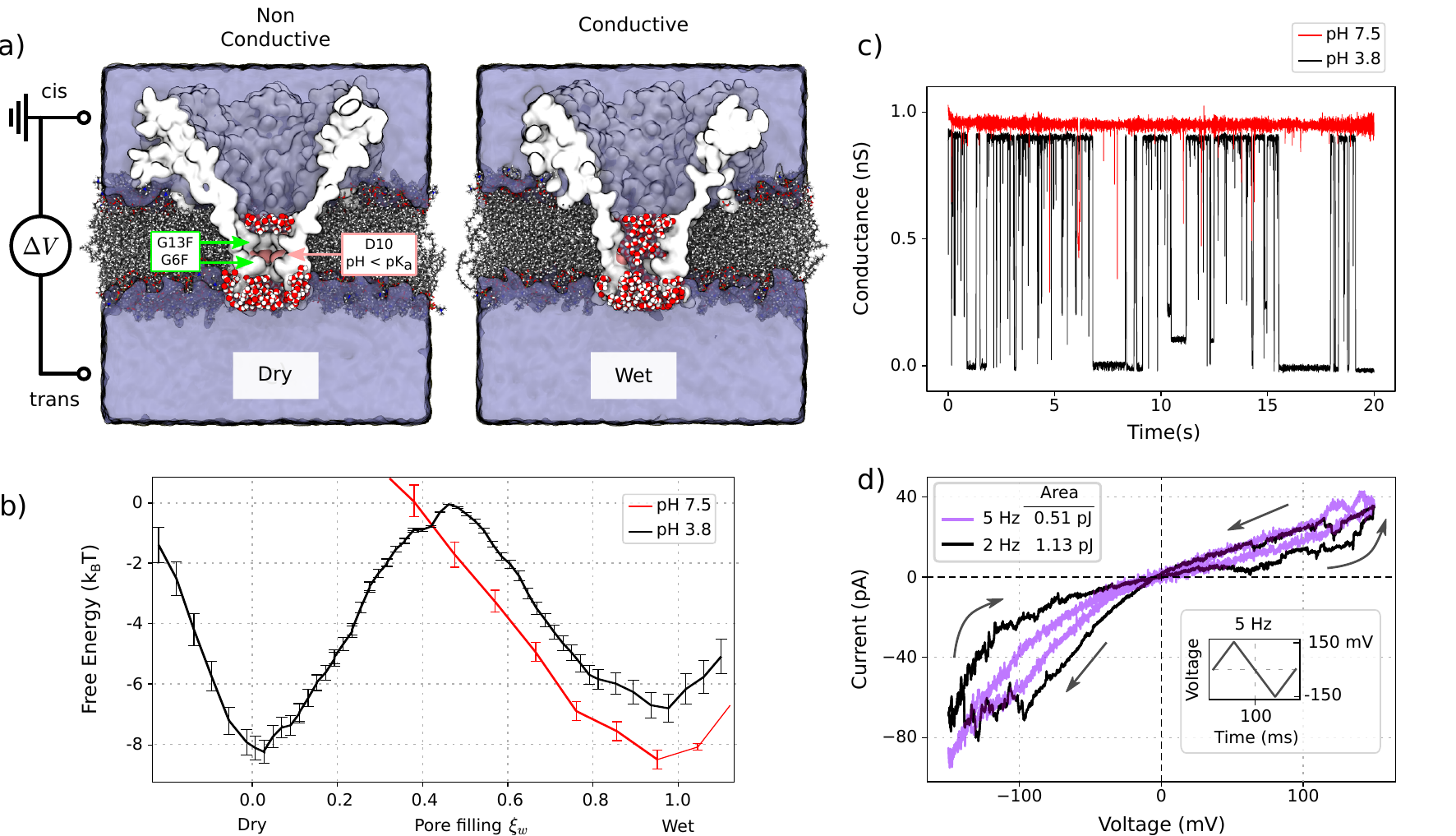}
    \caption{\textbf{Hydrophobically gated FraC nanopore}.
    \textbf{a)} MD system composed by the FraC nanopore 
    embedded in a lipid membrane and immersed in 1M KCl water solution.
    Ions are not show for sake of clarity.
    In pink are highlighted the acidic residues that are protonated at pH 3.8, e.g.,
    the aspartic acid D10.
    The system shows the FraC channel with a mutation on G13F and G6F,
    effectively creating a narrow hydrophobic region in the pore constriction.
    The water molecules inside the control box are displayed in classical VDW style.
    \textbf{b)} Pore filling free-energy profile, 
    computed by counting the number of water inside the control box
    (pore constriction), at pH 7 (red) and 3.8 (black).
    At neutral pH the system presents only one minimum, 
    corresponding to the wet state,
    while at low pH the system displays two minima, 
    with the dry state being the more probable.
    Error bars represents the standard error from block average with window length of 50 ps.
    \textbf{c)} Experimental current traces,
    single pore,
    measured at constant voltage $\Delta V = -50$~mV
    at different pH.
    Lowering the pH makes the channel gate,
    as the neutralisation of the charged residue is completed,
    and a hydrophobic region is developed.
    \textbf{d)} Experimental IV curve under a cycling applied voltage (period 0.5s).
    The plot is obtained by averaging the current at each voltage over 35 realisations of the same cycle, after the capacitance current was subtracted,
    see Supp.~Note~S4 and Supp.~Fig.~S13-19. 
    The systems clearly show a pinched hysteresis loop,
    the hallmark of memristors. The direction of the loop is that of a bipolar memristor
    \label{fig:fig3}}
\end{figure*}

To put to the test the above predictions, 
we engineered a biological nanopore 
-- the Fragaceatoxin C (FraC)-- to hydrophobically gate.
The wild type FraC is a biological toxin found in 
the sea anemone {\it Actinia fragacea}~\cite{Tanaka2015}, which has been recently used in  single-molecule nanopore sensing~\cite{Wloka2016}; its stability allows the pore to be easily engineered by introducing different point mutations in its constriction~\cite{lucas2021manipulation}.

In line with the design criteria of Fig.~\ref{fig:fig2} and supported by atomistic simulations,
we designed the double mutant {\it G6F,G13F-FraC} in which two hydrophobic residues are introduced in the narrowest region of the pore (the constriction), see Fig.~\ref{fig:fig3}a and Supp.~Fig.~S7.
A peculiar feature of the system is the presence of a titratable ring of aspartic acids D10
(pK$_\text{a}=4.5$) at the constriction center, between F6 and F13, that can be used to tune the wettability of the pore by changing the pH. Indeed, the protonation of D10, highlighted in Fig.~\ref{fig:fig3}a,
creates an uncharged region extending for ca. $1.2$~nm (3-4 aminoacidic rings)
that is mostly hydrophobic, allowing the formation of a stable vapour bubble inside the pore as demonstrated by the RMD simulations in Fig.~\ref{fig:fig3}b. 
Indeed the computed pore filling free-energy profile shows that, at pH 3.8, with D10 completely neutralised,
the system exhibits two free-energy minima, 
with the dry state being the most favourable one.
On the other hand, at pH 7.5 (red line), with charged D10,
the system displays a single free-energy minimum -- the wet (conductive) state.

The theoretical predictions are confirmed by single-pore electrophysiology measurements,
reported in Fig.~\ref{fig:fig3}b.
A random telegraph signal between two main conductance levels is observed at pH 3.8 (black line), which is not seen neither in the wild type nor in the mutated pore at pH 7.5 (red line).
In the light of the simulations in Fig.~\ref{fig:fig3}b, these results point to the presence of hydrophobic gating in the mutated pore.
In the absence of nanopores, our system displays leakage currents of the order of $0.05\pm 0.075$~nS, see Supp.~Fig.~S8, which makes it hard to distinguish whether the appearance of step-like levels at the lowest conductances ($<0.2$~nS) originates in conduction events through the membrane or in additional dynamics at the nanopore level, e.g., coming from the interplay between bubble formation and the flexible protein structure (e.g., elastocapillary effects~\cite{giacomello2020bubble, caprini2023metastable}), or interactions with contaminants in solution. 
In general, at low pH the nanopore acts as a bistable system, with the two main peaks -- associated to the fully open and closed pore -- accounting for c.a. 90\% of the average conductance of the system, as expected for the hydrophobically gated nanopore proposed by our model. While, in principle, subconductance levels cannot be excluded, they do not seem relevant for the HyMN concept and would introduce rather small variations in the closed current level.

Fig.~\ref{fig:fig3}d reports the average experimental IV curve, from which the capacitive current, e.g., due to the membrane, was subtracted out -- see Supp.~Note~S4 and Supp.~Figs.~S9-S13.
The system clearly shows a pinched hysteresis loop, characteristic of memristors.
The asymmetric response of the system, 
under opposite applied voltages, 
is likely to originate in the conical shape and non-uniform charge distribution in the FraC nanopore,
as previously reported for other asymmetric geometries~\cite{powell2011electric,Klesse2020}, 
see Supp.~Fig.~S14:
by using the same protocol as in  Fig.~\ref{fig:fig1}a to compute the effect of voltage on the simulated free-energy profile, we find that the intrinsic dipole of the FraC nanopore system leads to an asymmetric response under opposite voltages, consistent with experiments.
Differently from Fig.~\ref{fig:fig1}d, 
the IV curve self intersects at the origin, which is the signature of bipolar memristors (Supp.~Fig.~S6). 
Therefore, controlling the symmetry/asymmetry of the pore,
HyMN devices can be designed to behave either as unipolar or bipolar memristors. 
At this stage, the curve in Fig.~\ref{fig:fig3}d is due to the averaging of multiple single pore recordings, see Supp.~Fig.~S13. The same memristive behaviour could be obtained by having multiple pores in the same membrane patch, which would result in a device reacting to a single voltage pulse.

In summary, by engineering the wetting properties of a mutated FraC nanopore,
we demonstrated the potentiality of the proposed nanofluidic memristors,  
which exploit hydrophobic gating to induce memory as further demonstrated in the next section. 
HyMNs have the advantages of 
compactness, simplicity --having no moving parts nor allosteric gating mechanisms-- durability, and high reproducibility.
The mutagenesis approach can be easily extended to other well studied nanopores having different radii and lengths, such as $\alpha$-Hemolysin~\cite{deamer2016three}, Aerolysin~\cite{cressiot2019aerolysin}, CsgG~\cite{dimuccio2022geometrically}, or artificial de-novo $\beta$-barrel nanopores~\cite{shimizu2022novo}
that can have other dynamical characteristics.
Moreover, solid-state nanopores can be easily grafted with hydrophobic groups~\cite{powell2011electric,bandara2019chemically}
and, together with engineered biological nanopores,
can pave the way to the next generation of highly tunable nanofluidic memristors.

\subsection*{ Neuromorphic applications using HyMNs}
\begin{figure*}[h!]
    \centering
    \includegraphics[width=1\linewidth]{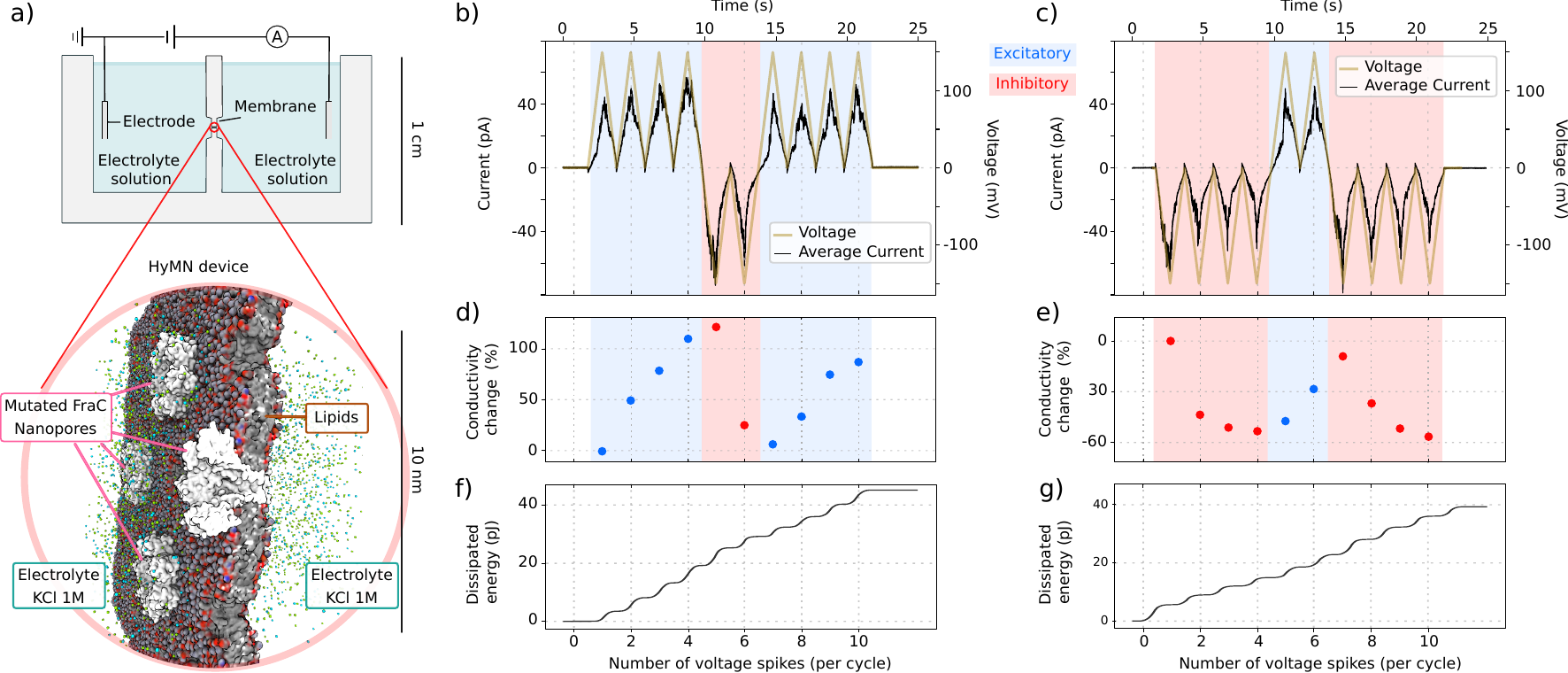}
    \caption{ \textbf{Learning and forgetting in a hydrophobically gated memristive nanopore (HyMN) device.}
    \textbf{a)} Experimental setup. 
    The HyMN, 
    here represented in the red inset, 
    is composed of multiple engineered FraC nanopores (less than 5, white pores), 
    immersed into a lipid bilayer (gray and red membrane) separating two electrolyte reservoirs (green and blue spheres, water not shown).
    The lipid membrane is suspended on a Delrin partition
    (Warner, USA) containing an aperture of approximately
    150 $\mu$m in diameter, see Materials and Methods for more details.
    The current that passes through a HyMN depends on the previous voltage applied at its terminals.
    \textbf{b)} The average of measured current (black line) through the HyMN subjected to a voltage signal (golden line) composed of 4 positive excitatory triangular waves, 2 negative inhibitory ones, and 4 positive ones. The current response increases during the excitatory pulses and decreases during the inhibitory ones.
    \textbf{c)} HyMN device response to the opposite protocol, starting with 4 inhibitory pulses, followed by 2 excitatory ones, and ending with 4 inhibitory pulses. Again, during the inhibitory stage the maximum current goes down, while it increases after excitatory pulses. For both panels b) and c) the capacitive current was subtracted; 
    the measured current is the average over 35 realizations. 
    \textbf{d-e)} Change in the average conductance (\%),
    computed with respect to the conductance during the first spike in b and c, respectively. The conductance is averaged over each pulse.
    Note that the pore is more conductive at negative voltages, 
    see Fig.~\ref{fig:fig3}d, a behaviour usually observed in biological nanopores and also reported for the FraC WT and other mutants~\cite{lucas2021manipulation}.
    \textbf{f-g)} Cumulative dissipated energy during the cycles reported in b and c, respectively. The dissipated energy is computed by integrating the mean electric power during the cycle, obtained by averaging the current trace over 35 realizations.}
    
    \label{fig:fig4}
\end{figure*}

Neuromorphic computing has garnered significant attention as it promises to transcend the capabilities of digital computers by emulating the complex behaviour of neurons. 
Here, we tested the potential of HyMNs in neuromorphic applications by experimentally realising a device that has a synapse-like learning-and-forgetting behaviour (Fig.~\ref{fig:fig4}a). We developed a FraC-based HyMN, constituted of few nanopores (less than 5) in a lipid membrane separating two electrolyte reservoirs. A sequence of voltage pulses is imposed and the current through the device is measured,
see Figure~\ref{fig:fig4}b-e. Successive pulses with $\Delta V>0$ cause a progressive increase in the current (learning), while for pulses with $\Delta V<0$ the current decreases (forgetting). In other words, the bipolar nature of the memristor allows for excitatory and inhibitory responses in the same device, allowing to program the device response depending on the history of received stimuli (memory). The possibility to reversibly
control the conductance of the device using excitatory and inhibitory pulses paves the way to the exploitation of HyMNs in more complex iontronic learning devices, e.g., for analog neural networks.

Although a direct comparison with different systems is difficult, we try to list advantages and disadvantages of HyMNs.
The energy consumption of our device during the synaptic events is on the order of some pJ, see Fig.~\ref{fig:fig4}f-g, on par with biological neurons which are estimated to consume between $0.01$ and $10$~pJ per voltage spike~\cite{Kuzum2013}, which is significantly less expensive than those produced by solid-state neurons, which in turn outperform  digital software-based ones~\cite{Poon2011}. For comparison, the 2D nanofluidic memristors of Ref.~\cite{robin2023}, requires order of nJ to perform similar tasks.
Similarly to other nanofluidic memristors~\cite{robin2023},  HyMNs can be designed to work as unipolar, Fig.~\ref{fig:fig1}, or bipolar Fig.~\ref{fig:fig3}, memristors, which can be engineered by considering their symmetry, see also Supp. Fig. S6.
As compared to voltage-gated ion channels, the presented HyMNs have no moving parts, and hence are more easily tunable and robust, as specific mutations of the pore lumen have a predictable effect on hydrophobic gating, as demonstrated in this work;
differently, mutations on voltage-gated ion channels have more complicated implications on the protein structure and on the allosteric gating mechanisms~\cite{Bassetto2023} and, hence, are harder to engineer. 
Similarly to ion channels, HyMNs can be tuned to work from stochastic memristors all the way to deterministic ones (multiple channels, as in neurons). 
Although the HyMN concept is rather general, the FraC realisation also has some drawbacks; the pore is an alpha-helix assembly, which makes it more flexible and potentially less robust than $\beta$ barrel pores or solid state ones. Other  candidate channel are currently being explored to  further expand HyMN capabilities.
Our findings showcase the potential of HyMNs as flexible building blocks of nanofluidics neuromorphic computing.

%%%%%%%%%%%%%%%%%%%%%%%%%%%%%%%%%
\section*{Discussion}
In this work, we propose and demonstrate a hydrophobically gated memristive nanopore (HyMN). Molecular dynamics simulations revealed the microscopic mechanism at the heart of the memristive behaviour, i.e., memory by electrowetting. Guided by the molecular dynamics results, we propose design criteria to narrow the parameter space where HyMNs can be found, pointing towards biological nanopores as promising candidates owing to their size and the possibility to carefully control their hydrophobicity by point mutations.
We tested our prediction by engineering a mutant of the biological FraC nanopore to have a hydrophobic constriction.
Molecular dynamics simulations demonstrated that it displays hydrophobic gating at low pH. 
Electrophysiological experiments confirmed this microscopic insight, 
showing a random telegraph signal only at low pH and displaying the hysteresis 
loop in the IV curve, which is a signature of memristors. 
A HyMN-based device was successfully built and tested, showing synaptic capabilities, 
harnessing the power of hydrophobic gating and electrowetting to learn and forget.
We show that engineered biological nanopores thus can serve as HyMNs, with important strengths: 
they are energy efficient, nanometer-sized, have no moving parts, are highly reproducible and economical, and advanced technologies are available to fine tune their properties~\cite{deamer2016three,shimizu2022novo}.

The computational capabilities of the brain have initiated the era of artificial intelligence, 
which in turn  calls for suitable neuromorphic computing architectures, 
that should be durable and sustainable.
The most advanced technologies so far have employed semiconductors, but nanofluidic memristors are making way. The proposed HyMN concept brings back to the original archetype from which this journey started, i.e., ion channels which confer to neurons their computational capabilities. Could the considerable simplification of hydrophobic gating together with the current capabilities of molecular biology bring about a revolution in the field?

\section*{Methods}

\subsection*{Molecular dynamics setup}
We used molecular dynamics simulations to extract the free energy profile,
diffusivity, and their dependence with voltage. 
These simulations were done using the molecular dynamics package LAMMPS~\cite{LAMMPS} for the model pore
and NAMD~\cite{phillips2005scalable} for the FraC.

\textbf{Model Nanopore.}
We construct a membrane out of a slab of fixed atoms in a FCC arrangement, 
with lattice spacing 0.35 $nm$, from which a nanopore was excavated.
Water (SPC/E~\cite{berendsen1987missing}) was placed on both sides of the slab.
The interaction of the water with the surface of the material was tuned~\cite{camisasca2020gas} so that the contact angle of water droplet on top of the material would be $104^\circ$.
The nanopore has a nominal diameter of 1.04 $nm$ and a nominal length of 2.8 $nm$. 
At each end of the water reservoirs, a thin slab of hydrophilic material is present,
which is used as a piston to control the pressure of the system~\cite{marchio2018}.
The NVT ensemble was sampled using a Nos\'e--Hoover chains thermostat \cite{martyna1992} 
at 310$^\circ$ K with a chain length of 3.

\textbf{FraC Nanopore.} 
The starting PDB structure of the heptamer nanopore
was taken from the previous work~\cite{Huang2019}.
The missing N-terminals (sequence ASAD) were modeled by AlphaFold;
the modeled sequence was ASADVAGAVIDGAGLG, generating an alpha-helix (best prediction) 
that was then aligned and merged to each of the seven chain of the starting FraC structure.
After, the residues G6 and G13 were mutated in phenylalanine with VMD~\cite{vmd} Mutator Plugin.
The resulting structure is minimised for 1000 step of gradient descent in vacuum.
From the mutated pore, two systems were built: one at pH 7 and another at pH 3.8.
The protonation state of each tritable residue was assessed by computing the $\mathrm{pK_a}$ 
with 
PROPKA~\cite{olsson2011propka3}. 
In particular, it resulted that the D10 residues, at the centre of the hydrophobic region,
have an average pK$_\text{a}= 4.66 \pm 0.06$; so, they are completely protonated at pH 3.8.
The complete list of pK$_\text{a}$ and protonation state for the other amino acids can be found in Supp.~Fig.~S7.
The two systems were embedded in a lipid bilayer and immersed into a 1M KCl solution, 
using the VMD's Membrane Builder, Solvate and Autoionize Plugins.
The final system consists of about 215,000 atoms
($\sim$450 molecules of 1-palmitoyl-2-oleoyl-sn-glycero-3-phosphocholine (POPC),
50,000 water molecules, 850 potassium and 900 chloride ions)
and the simulation box had dimensions (142 × 142 × 130)~\AA$^3$.
For the MD simulations, we used the ff15ipq force field~\cite{debiec2016further} for the protein, 
the Lipid17 force-field \cite{dickson2014lipid14} for the phospholipids 
and the SPC/E water model~\cite{takemura2012water}.
The systems were thermalised and equilibrated for 10 ns,
following the procedure illustrated in previous works~\cite{dimuccio2019insights}.

\subsection*{Restrained molecular dynamics}
We use Restrained Molecular Dynamics (RMD)~\cite{maragliano2006temperature} 
to compute the free energy as a function of the pore filling.
This is done by adding a harmonic restraint to the original Hamiltonian of the system,
\begin{equation}\label{eq:restraint}
    H_N(\boldsymbol{r},\boldsymbol{p})=H_0(\boldsymbol{r},\boldsymbol{p})+\frac{k}{2}\left(N-\tilde{N}(\boldsymbol{r})\right)^2\,,
\end{equation}
where $\boldsymbol{r}$ and $\boldsymbol{p}$ are the positions and momenta of all the atoms, respectively, $H_0$ is the unrestrained Hamiltonian,
$k$ is a harmonic constant which was set to $1$\,kcal/mol, 
$N$ is the desired number of water molecules in a box centred around the nanopore, 
and $\tilde{N}$ is the related counter, 
counting the number of water molecules in the system.
We used the same protocol for counting the number of water molecules as described in previous work \cite{Paulojcp2023}.
A fermian distribution with Fermi parameter equal to 3 \AA 
is used to smooth the borders of the box and make the 
collective variable continuous. 
For the FraC nanopore, the protocol is implemented in NAMD by using the 
Volumetric map-based variables of the Colvars Module~\cite{fiorin2013using}.
For the FraC nanopore, the centre and size of the counting box was set 
equal to the centre and the size of the F6 and F13 hydrophobically mutated rings.
The water molecules affected by the counting box are represented in VDW spheres in Fig. 3 of the main manuscript.
Each filling state, corresponding to each point of Fig.~3c,
was sampled for 2 ns.
Each trajectory is saved every 20ps,
while the number of water molecules inside the box is saved every 1 ps.

\subsection*{Experimental setup}
\textbf{Chemicals.}
Potassium chloride, sodium chloride, imidazole, urea, Citric acid, N,N-dimethyldodecylamine N-oxide (LDAO), chloroform, n-decane and LB medium were purchased from Sigma-Aldrich. 1,2-Diphytanol-sn-glycero-3-phophocholine (DPhPC) lipids and sphingomyelin were obtained from Avanti. Ampicillin and isopropyl-$\beta$-D-1-thiogalactopyranoside (IPTG) were purchased from Fisher Scientific. 

\textbf{Mutant FraC expression.}

The {\it G13F-FraC} and {\it G6F,G13F-FraC} variants were prepared from the wild-type FraC gene ({\it wtFraC}) \cite{Lucas2021,Huang2019} by site-directed mutations were introduced using a megaprimer-based approach. In short, the T7 terminator primer (5’ CCGCTGAGCAATAACTAGC3’) and {\it G13F-FraC} mutant primer (5’GACGGTGCGTTCCTGGGCTTTGAC3’) were used to amplify a ‘megaprimer’ using the {\it wtFraC} plasmid  as template using a Phire Hot Start II Polymerase using the following conditions: denaturation at 98°C for 3 min, followed by 35 cycles of 98°C for 10 s, 55°C for 30 s, and 72 °C for 30 s, and a final extension cycle of 72°C for 5 min. After the PCR reaction, the parental DNA template (616 bp) was purified using GeneJET PCR Purification Kit (Fisher Scientific). This resulting DNA was then utilized as a ‘megaprimer’ to amplify the plasmid containing the {\it wtFraC} gene and generate {\it G13F-FraC} (same conditions as described above except for the 1 min extension time). 1.0 $\mu$L of the PCR mixture was then incorporated into 50 $\mu$L E. cloni® 10G (Lucigen) competent cells by electroporation. Transformants were grown overnight at 37°C on LB agar plates supplemented with ampicillin (100 $\mu$g/mL). The plasmid was extracted with GeneJET Extraction Kit (Fisher Scientific) and the identity of the clones was confirmed by sequencing. The {\it G6F,G13F-FraC} variant was obtained in a similar fashon using {\it G13F-FraC} plasmid as a template.

\textbf{FraC monomer expression and purification.}
A pT7-SC1 plasmid containing the {\it G6F,G13F-FraC} gene was transformed into BL21(DE3) cells. The transformed cells were grown overnight at 37$^\circ$~C on LB agar plates supplemented with 
1\% glucose and 100 $\mu$g/ml ampicillin. On the next day, picking the single colony and resuspended to grow in 10 mL LB medium at 37 ºC overnight. Grown LB culture was transferred into 1 L LB medium, supplemented with 100 mg/L ampicillin and grown under constant shaking at 37 °C until the OD600 reached a value of 0.8-1.0. At this point, 0.5 mM IPTG was added for induction and growth continued overnight at 25 oC. Afterwards, the cells were pelleted by centrifugation at 3500 rpm for 15 minutes. 50 mL lysis buffer, containing 150 mM NaCl, 2 M Urea, 20 mM imidazole and 15 mM Tris buffered to pH 7.5. The mixture was sonicated using a Branson Sonifier 450 (2 minutes, duty cycle 30\%, output control 3) to ensure full disruption of the cells. The lysate was pelleted by centrifugation at 10000 rpm for 30 minutes and the supernatant is carefully transferred to a fresh falcon tube. Meanwhile, 300 $\mu$l of Ni-NTA bead solution is washed with wash buffer, containing 150 mM NaCl, 20 mM imidazole and 15 mM Tris buffered to pH 7.5. The beads are added to the supernatant and incubated at room temperature for 30 minutes under constant rotation. Afterwards, the solution is loaded on a pre-washed Micro Bio-Spin column (Bio-Rad) and washed with 50 ml of wash buffer. The bound protein is eluted in fractions of 200 $\mu$l of elution buffer (150 mM NaCl, 300 mM imidazole, 15mM Tris buffered at pH 7.5). The presence of FraC monomers was detected using SDS-PAGE.

\textbf{Sphingomyelin-DPhPC liposomes preparation.}
An equal mixture of 25 mg sphingomyelin and 25 mg DPhPC was dissolved in 10 mL pentane containing 10 v/v\% chloroform. A film was formed on the side of the flask through evaporating by nitrogen under constant rotation. The resulting film was dissolved in 10 mL of 150 mM NaCl and 15 mM Tris buffered to pH 7.5. The resulting liposome solution (5 mg/mL) was frozen (-20 °C) and thawed multiple times.

\textbf{FraC oligomerisation.}
Liposome were added to FraC monomers in a 10:1 (lipid: protein) mass ratio at 37 oC for 30 minutes. Afterwards, the liposomes were solubilised by the addition of 0.6\% v/v\% LDAO. Subsequently, the sample was diluted 10 times with 150 mM NaCl buffered at pH 7.5 using 15 mM Tris supplemented with 0.02\% DDM. Meanwhile, 200 $\mu$l of Ni-NTA was added and incubated for 1 h at 25 ºC under constant rotation to purify the FraC oligomer. The mixture was then loaded on a Micro Bio-Spin column and extensively washed with 20 mL of wash buffer supplemented with 0.02\% DDM. FraC oligomers were eluted in 200 $\mu$l elution buffer containing 1 M imidazole, 150 mM NaCl and 15 mM Tris buffered to pH 7.5 supplemented with 0.02\% DDM. Oligomers can be stored at 4 oC for several weeks.

\textbf{Planar lipid bilayer electrophysiological recordings.}
Two fluidic compartments are separated into cis and trans compartments by a Delrin partition (Warner, USA) containing an aperture of approximately 150 $\mu$m in diameter. An Ag/AgCl electrode is placed in each compartment as to make contact with the buffer solution. Planar lipid bilayers were formed in single-channel trials using the standard methods as described previously. In brief, the 150 $\mu$m aperture in a Delrin partition was pre-painted with 0.5 $\mu$L of 40 mg/mL DPhPC solution in n-decane prior to loading the buffer containing 1 M KCl buffered at pH 3.8 using 50 mM citric acid with bis-tris-propane, and then painted with 0.2 $\mu$L of 40 mg/mL of DPhPC for the bilayer formation. {\it G6F,G13F-FraC} oligomers were added to the cis compartment which was connected with the ground. Presence of a single channel were confirmed by the IV characteristics of the pore. After buffer exchange in cis compartment to remove additional oligomers, 
gating data were collected by applying different potential sequence protocols.

\textbf{Data acquisition.}
The ionic current was recorded using a Digidata 1550B (Molecular Devices) connected to an Axopatch 200B amplifier (Molecular Devices). Data is recorded with a sampling frequency of 10 kHz or 20 kHz. 
\section*{Data availability}
The simulation and experimental data used in this study are available in the Zenodo database at 10.5281/zenodo.8018059.

%\bibliography{biblio.bib}
\bibliography{mybib}

\section*{Acknowledgements}
This research is part of a project that has received funding from the European Research Council (ERC) 
under the European Union's Horizon 2020 research and innovation programme (grant agreement No. 803213)(A. Giac.). 
The authors acknowledge PRACE for awarding us access to Marconi100 at CINECA.
The authors acknowledge CSCS for awarding us access to Piz Daint (project id s1178).
\section*{Authors contributions}
A.Giac. and M.C. conceived the project.
G.P. carried out the theoretical analysis and the molecular dynamics simulations on the model nanopore,
with inputs from A.Gub. and A.Giac.
K.S. performed the experiments under the supervision of G.M. and J.G.
G.D.M. and B.M. carried out the molecular dynamics simulations on the FraC channel.
G.P and G.D.M. wrote the article.
G.P and G.D.M. analysed the experimental data.
All authors discussed the results and reviewed the final manuscript.
\section*{Conflicts of interest}
The authors declare no competing interests.

\end{document}

\maketitle
\newpage

\noindent {\bf Contents:}
\\
\\
\footnotesize{
%\begin{tabular}{ll}
\begin{tabular}{p{15cm} p{4 cm}}

Supplementary Note S1: Effect of transmembrane voltage on free energy.            
& p. S-3  \\
Supplementary Note S2: Rate calculations for the wetting and drying events.       
& p. S-6  \\

Supplementary Note S3: Macroscopic model based on classical nucleation.           
& p. S-7 \\
Supplementary Note S4: Analysing of the voltage cycles of the FraC nanopore.      
& p. S-9 \\

\\
Supplementary Figure S1: Free-energy and diffusivity of pore filling.              
& p. S-10 \\
Supplementary Figure S2: Values of the first order coefficient, $P_1$, and the second order coefficient, $P_2$.       
& p. S-11 \\
Supplementary Figure S3: Estimation of the wetting/drying rates.                   
& p. S-12 \\
Supplementary Figure S4: Collective pore wetting.                                  
& p. S-13 \\
Supplementary Figure S5: Dependence of the hysteresis area with the applied voltage.
& p. S-14 \\

Supplementary Figure S6: Type of memristor behaviour.                       
& p. S-15  \\
Supplementary Figure S7: Molecular model of FraC nanopore.
& p. S-16 \\
\tr{Supplementary Figure S8: Single pore current traces at constant applied voltage and pH 3.8.}
& p. S-17 \\
\tr{Supplementary Figure S9: Experimental current signal of a voltage cycle of 5Hz through FraC nanopore at pH 3.8.}
& p. S-18 \\
\tr{Supplementary Figure S10: Experimental current signal of a voltage cycle of 2Hz through FraC nanopore at pH 3.8.}
& p. S-19 \\
Supplementary Figure S11: Removing the capacitive current at 5Hz and pH 3.8. 
& p. S-20 \\
\tr{Supplementary Figure S12: Removing the capacitive current at 2Hz and pH 3.8.}
& p. S-21 \\
\tr{Supplementary Figure S13: From single channel trace to average memristive behaviour.}
& p. S-22 \\
Supplementary Figure S14: Effect of the voltage on the free energy profile of FraC nanopore.
& p. S-23 \\

Additional References:                                                                 
& p. S-23 \\
\end{tabular}
}

%%%%%%%%%%%%%%%%%%%%%%%%%%%%%%%%%%%%%%%%%%%%%%%%%%%%%%%%%%%%%%%%%%%%%%%%%%%%%%%%%%%%%%%%%%%%%%%%

\newpage

%%%%%%%%%%%%%%%%%%%%%%%%

\newpage
\section*{Supplementary Note 1.\\
\large{Effect of transmembrane voltage on free energy}}

In this section we explore the effect that a transmembrane voltage has on the free energy profile associated with a given colective variable, which in our case is the probability of having the pore filled.
Here the main assumption is that the system is at the equilibrium after applying the voltage.
Consider a system described by the vector $\boldsymbol{x}$ of positions and momenta of its $N$ particles
$\boldsymbol{x}=(\boldsymbol{r_1},\boldsymbol{p_1},\dots,\boldsymbol{r_N},\boldsymbol{p_N})$,
with an Hamiltonian made by a ‘‘zero voltage'' part $H_0$ and the contribution of the external voltage $V$
\begin{equation}
H(\boldsymbol{x},V)=H_0(\boldsymbol{x})-Q_z(\boldsymbol{x})V\;,
\end{equation}
with $Q$ displacement charge
\begin{equation}
Q(\boldsymbol{x})=\frac{\pi(\boldsymbol{x})}{L}\;,
\end{equation}
where 
\begin{equation}
\pi(\boldsymbol{x})=\sum\limits_{i=1}^N q_i\boldsymbol{r}_i\cdot\boldsymbol{\hat{n}}
\end{equation}
is the total dipole moment of the system in the direction in which the voltage is applied $\boldsymbol{\hat{n}}$, and $L$ is the extension of the system in the same direction.

The probability density function of the system in the canonical ensemble is then
\begin{equation}
\rho(\boldsymbol{x},V)=\frac{1}{Z(V)}e^{-\beta H(\boldsymbol{x},V)}\;,
\end{equation}
with $Z$ partition function
\begin{equation}
Z(V)=\int e^{-\beta H(\boldsymbol{x},V)}\mathrm{d}\boldsymbol{x}\;.
\end{equation}

Choosing a collective variable $\theta$ which gives a coarser description of the system, the probability density of that variable is 
\begin{equation}
P(\theta,V)=\int \delta\left(\theta-\hat{\theta}(\boldsymbol{x})\right)\rho(\boldsymbol{x},V)\mathrm{d}\boldsymbol{x}=\frac{Z_\theta(\theta,V)}{Z(V)}\;,
\end{equation}
where $\hat{\theta}(\boldsymbol{x})$ is the function which associates the microscopic state to a value of the collective variable, and $Z_\theta$ is the partition function restrained to a given value of the collective variable 
\begin{equation}
Z_\theta(\theta,V)=\int \delta\left(\theta-\hat{\theta}(\boldsymbol{x})\right)e^{-\beta H(\boldsymbol{x},V)}\mathrm{d}\boldsymbol{x}\;.
\end{equation}
The associated free energy is
\begin{equation}
F(\theta,V)=-\beta\log P(\theta,V)\;,
\end{equation}
and its variation with respect to voltage
\begin{equation}
\frac{\partial F(\theta,V)}{\partial V}=-\frac{\beta}{P(\theta,V)}\frac{\partial P(\theta,V)}{\partial V}\;.
\end{equation}
The derivative of the coarse grained probability with respect to voltage is
\begin{equation}
\frac{\partial P(\theta,V)}{\partial V}=\int \delta\left(\theta-\hat{\theta}(\boldsymbol{x})\right)\frac{\partial \rho(\boldsymbol{x},V)}{\partial V}\mathrm{d}\boldsymbol{x}\;,
\end{equation}
with
\begin{equation}
\frac{\partial \rho(\boldsymbol{x},V)}{\partial V}=\left(\frac{\beta Q(\boldsymbol{x})}{Z(V)}-\frac{\mathrm{d}Z}{\mathrm{d}V}\frac{1}{Z(V)^2}\right)e^{-\beta H(\boldsymbol{x},V)}
\end{equation}
and 
\begin{equation}
\frac{\mathrm{d} Z(V)}{\mathrm{d}V}=\int \beta Q(\boldsymbol{x})e^{-\beta H(\boldsymbol{x},V)}\mathrm{d}\boldsymbol{x}=\beta Z(V)\left\langle Q\right\rangle_V\;.
\end{equation}
Hence, we obtain an expression for the derivative of the microscopic probability density with respect to voltage
\begin{equation}\label{eq:pdf_derivative}
\frac{\partial \rho(\boldsymbol{x},V)}{\partial V}=\frac{\beta}{Z(V)}\left( Q(\boldsymbol{x})-\left\langle Q\right\rangle_V\right)e^{-\beta H(\boldsymbol{x},V)}\;,
\end{equation}
which can be used to compute the same for the coarse grained probability density
\begin{equation}
\frac{\partial P(\theta,V)}{\partial V}=\int \delta\left(\theta-\hat{\theta}(\boldsymbol{x})\right)\frac{\beta}{Z(V)}\left( Q(\boldsymbol{x})-\left\langle Q\right\rangle_V\right)e^{-\beta H(\boldsymbol{x},V)}\mathrm{d}\boldsymbol{x}\;,
\end{equation}
simplifying
\begin{equation}
\frac{\partial P(\theta,V)}{\partial V}=P(\theta)\left(\left\langle Q\right\rangle_{\theta,V}-\left\langle Q\right\rangle_V\right)\;.
\end{equation}

We finally obtain an expression for the first order derivative of the free energy with respect to voltage
\begin{equation}
\frac{\partial F(\theta,V)}{\partial V}=-\beta\left(\left\langle Q\right\rangle_{\theta,V}-\left\langle Q\right\rangle_V\right)
\end{equation}

To obtain the second order derivative
\begin{equation}
\frac{\partial F(\theta,V)^2}{\partial^2 V}=-\beta\left(\frac{\partial\left\langle Q\right\rangle_{\theta,V}}{\partial V}-\frac{\partial\left\langle Q\right\rangle_V}{\partial V}\right)\;,
\end{equation}
we need to compute the derivative of the expected averages. Consider the expected average of a generic observable $A$
\begin{equation}
\frac{\partial \left\langle A\right\rangle_V}{\partial V}=\int A(\boldsymbol{x})\frac{\partial\rho(\boldsymbol{x},V)}{\partial V}\mathrm{d}\boldsymbol{x}\;,
\end{equation}
we can use Eq\eqref{eq:pdf_derivative}
\begin{equation}
\frac{\partial \left\langle A\right\rangle_V}{\partial V}=\int \frac{\beta A(\boldsymbol{x})}{Z(V)}\left( Q(\boldsymbol{x})-\left\langle Q\right\rangle_V\right)e^{-\beta H(\boldsymbol{x},V)}\mathrm{d}\boldsymbol{x}\;,
\end{equation}
and finally we get
\begin{equation}
\frac{\partial \left\langle A\right\rangle_V}{\partial V}=\beta\left(\left\langle AQ\right\rangle_V-\left\langle A\right\rangle_V\left\langle Q\right\rangle_V\right)\;.
\end{equation}

Now we can compute the second order variation of the free energy with voltage
\begin{equation}
\frac{\partial F(\theta,V)^2}{\partial^2 V}=-\beta^2\left(\left\langle Q^2\right\rangle_{\theta,V}-\left\langle Q\right\rangle^2_{\theta,V}-\left\langle Q^2\right\rangle_{V}+\left\langle Q\right\rangle^2_{V}\right)\;.
\end{equation}

\begin{figure}[h!]
    \centering
    \includegraphics[width=\linewidth]{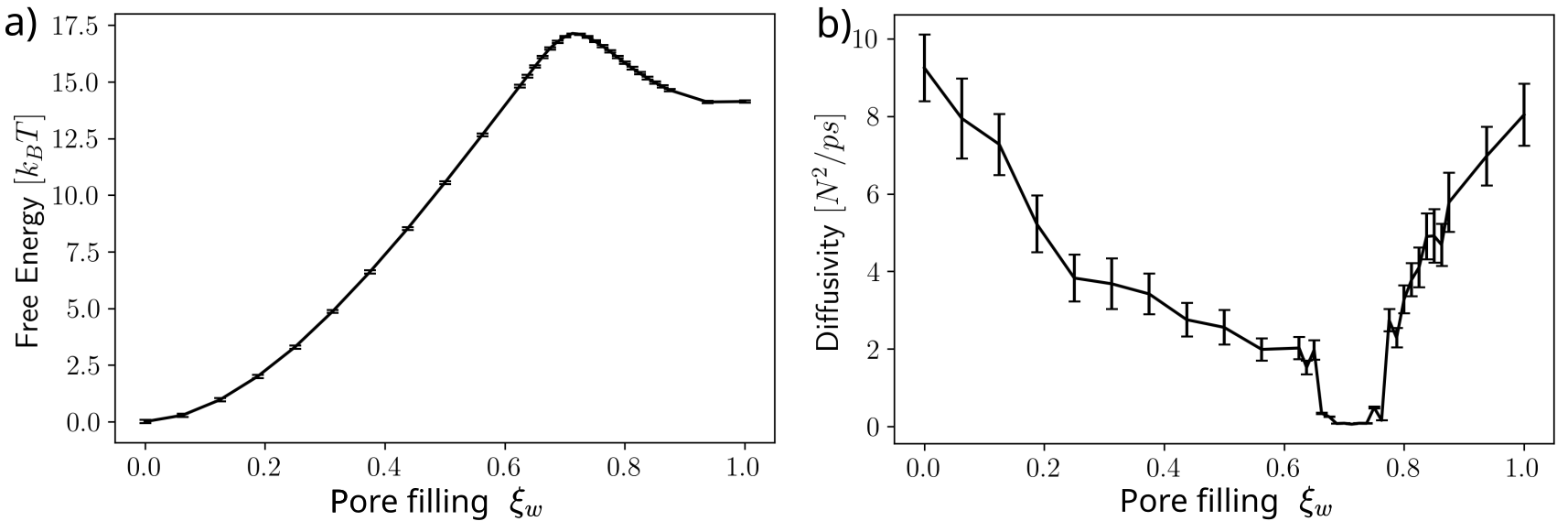}
    \caption{\textbf{Free energy and diffusivity as a function of water filling. }
    We used Restrained Molecular Dynamics simulations to compute the free energy (panel a) associated with the level of water filling the model pore of Fig 1,
    and the corresponding state diffusivity (panel b),
    following the protocol described before in a previous work~\cite{Paulojcp2023} and in the methods section. The error on the estimation of the free energy is estimated by separating the data in 100 blocks, integrating each independently and then averaging the resulting profiles. The error on the estimation of the diffusivity is described before in a previous work~\cite{Paulojcp2023}.
    As mentioned in the main text, the pore has a stable (dry, $\xi_w$= 0) and a metastable state (wet, $\xi_w$= 1) when no voltage is applied.
    The diffusivity of the wetting variable depends on the value of the wetting variable, 
    and changes more than one order of magnitude from its maximum value to its minimum value. 
    The diffusivity of the wetting variable is necessary to compute the correct wetting and drying times and their dependence on voltage. 
    We consider that this diffusivity does not depend on the voltage, as we have shown previously that it does not depend on the applied pressure~\cite{Paulojcp2023}.
    Here we report the water filling, which we compute as the number of water molecules inside the pore divided by the number of water molecules in the wet state. 
    These particular results we previously reported in another work \cite{Paulojcp2023}.
    \label{fig:freedifusivity}
    }
\end{figure}

\begin{figure}[h!]
    \centering
    \includegraphics[width=\linewidth]{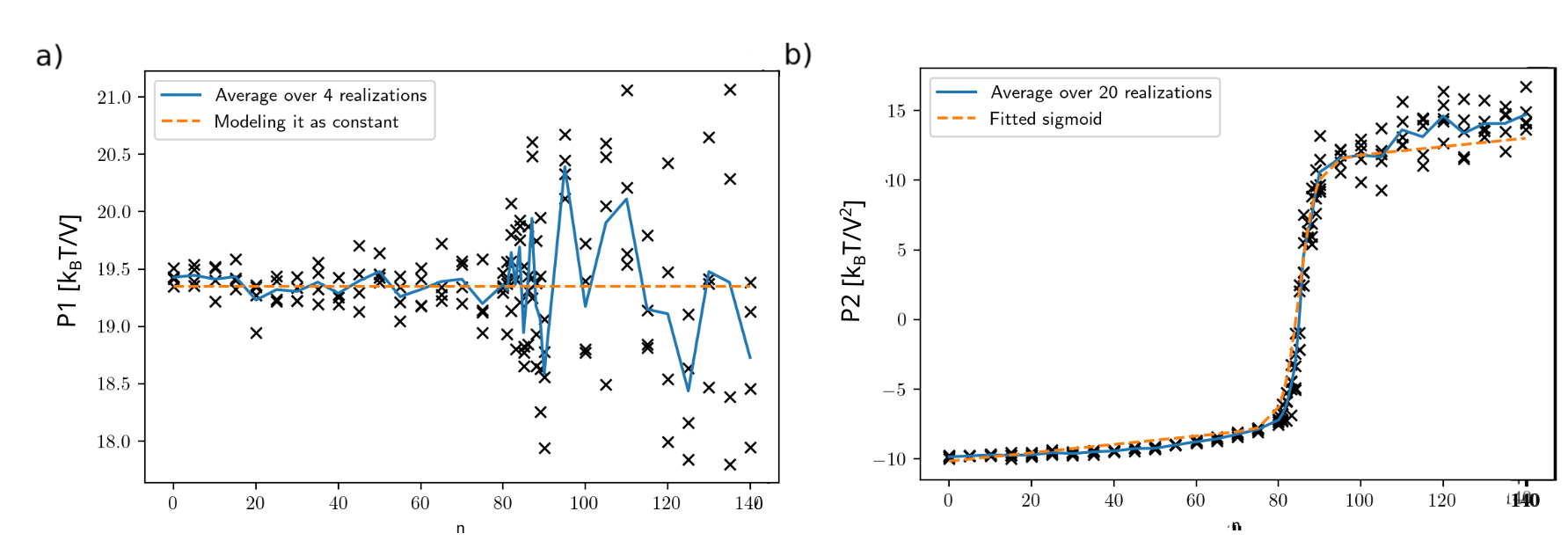}
    \caption{\textbf{Values of the first order coefficient, $P_1$, and the second order coefficient, $P_2$.} 
    During the restrained molecular dynamics simulations, $P_1$ and $P_2$, the first and second order coefficients on the expansion of the free energy discussed in Supp. Note 1, were calculated at different pore fillings, n. The pore filing in this figure is related to the number of water molecules inside the control box used in the RMD simulations. Because the pore is symmetric and uncharged we model $P_1$ as constant, panel (a), not having any effect on the free energy, as the free energy is always defined up to a constant. $P_2$ can be fitted by a sigmoid function plus a straight line, panel (b). Because the free energy is always defined up to a constant we set the $P_2$ equal to zero at the maximum of the free energy profile, n=90, corresponding to $\xi_w$= 0.7 in figure S1. }
    \label{fig:displacementcharge}
\end{figure}

The free energy dependence on the voltage can be then approximated as:

\begin{equation}
    F(\theta,V) = F(\theta) + P_1 V + P_2 V^2 \;,
\end{equation}
where $P_1 = -\beta \left\langle Q\right\rangle_{\theta}$ and $P_2=-\beta^2(\left\langle Q^2\right\rangle_{\theta}-\left\langle Q\right\rangle^2_{\theta})$.
During the restrained molecular dynamics simulations used to compute the profiles of Supp.~Fig.~\ref{fig:freedifusivity} we $P_1$ and $P_2$ for different filling levels, Supp.~Fig.~\ref{fig:displacementcharge}.

Because our pore is i) symmetric, ii) uncharged and iii) the displacement charge enters as a factor proportional to voltage, the displacement charge must not depend on the filling of the pore,
as positive and negative voltages have the same effect on wetting and drying.
Hence, we considered the displacement charge constant and independent of the filling level,
i.e. we can discard the first order dependence on the applied voltage.
The variance of the displacement sharply increases when the pore goes from dry state to the wet state, 
with the switching happening close at the same filling level where the free energy barrier occurs. 
We fit this behaviour with a sharp sigmoidal function. 
The variance of the dipole has the units of a capacitance, 
and the term added to the free energy could be related to the energy stored in the pore if it were a capacitor.
The dry pore has the same shape of the wet pore, but it is filled with a lower dielectric (air vs water) and, so,
the capacitance of the dry pore is lower than that of the wet pore.
This means that long and narrow pore will feel less the effect of the electric field,
with respect to shorter and wider ones. 
We underline that the difference in the capacitive energy is not due to the shape of the whole pore,
but only of the hydrophobic region where the bubble is observed.
The sole dependence of the wetting probability on the square of the voltage has been reported in 
works by different authors~\cite{Trick2017}.

%%%%%%%%%%%%%%%%%%%%%%%%%%%%%%%%%%%%%%%%%%%%%%%%%%%%%%%%%%%%%%%%%%%%%%%%%%%%%%%%%%%%%%%%%%%%%%%%%%%%%%%%%%%%%%%%%
\newpage
\section*{Supplementary Note 2.\\
\large{Rate calculations for the wetting and drying events.}}

\begin{figure}[h!]
    \centering
    \includegraphics[width=\linewidth]{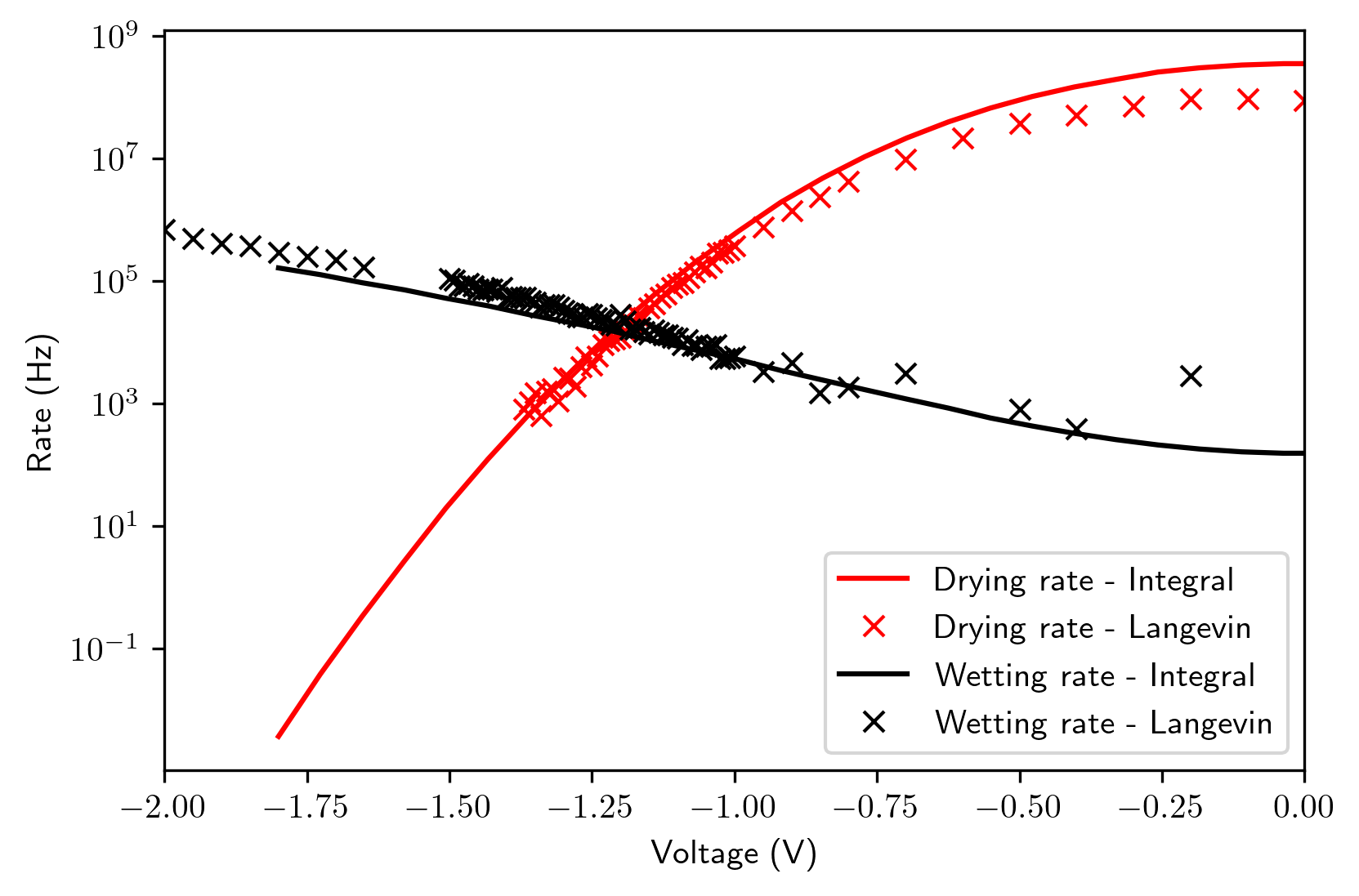}
    \caption{ \textbf{Estimation of the wetting/drying rates.} We estimate the wetting(red)/drying(black) rates using two different methods. The integral method involves solving Eq \ref{eq:ratetheory}, and is the one use throughout this work. The other is simulating the evolution of the Langevin equation, as done in previous work \cite{Paulojcp2023}. Both techniques give similar results, and due to the fact that the integral method is easier to compute, we use these results throughout the work. The point where the rates intersect is the value of $V_c$ reported in figure 1 of the main text. The value of $V_c$ is ca. 1.2 V.}
    \label{fig:rates}
\end{figure}

To compute the wetting/drying times shown in Fig.~1c of the main text we used the rate theory~\cite{Hnggi1990,zhu2012theory}
fed with the free energy profile $F(N,V)$, where N is the pore filling and V the applied voltage, and the state dependent diffusivity $D(N)$
\begin{equation}
    t_{rt} = \int_{N_a}^{N_b}\frac{e^{\frac{F(N,V)}{k_BT}}}{D(N)}dN \int_{N_a}^{N_b}e^{\frac{-F(N,V)}{k_BT}}\phi(N) dN,
    \label{eq:ratetheory}
\end{equation}
with $t_{rt}$ being the mean transition time and $N_a$ and $N_b$  the filling levels of the initial and final states and where $\phi(N)$, the committor, is given by
\begin{equation}
    \phi(N) = \frac{\int_{N_a}^{N}\frac{e^{\frac{F(N',V)}{k_BT}}}{D(N')}dN'}{\int_{N_a}^{N_b}\frac{e^{\frac{F(N',V)}{k_BT}}}{D(N')}dN'},
    \label{eq:comittor}
\end{equation}
Using $N_a$ as the dry state and $N_b$ as the filled state yields the wetting time;
conversely, using $N_a$ as the wet state and $N_b$ as the dry state yields the drying time. 
Throughout the work, the wetting and drying rates are always computed using Eq~\ref{eq:ratetheory}.
The rates could also be computed using Langevin dynamics as done in previous works \cite{Paulojcp2023}. 
The (overdamped) Langevin equation associated with the pore filling $N$ and $V$ is
\begin{equation}
    \label{eq:langevin}
    \frac{\mathrm{d}N(V)}{\mathrm{d}t}=-\frac{D(N)}{k_B T} \frac{\mathrm{d}F(V)}{\mathrm{d}N}+\frac{\mathrm{d}D}{\mathrm{d}N}+\eta(N)\xi(t)\,
\end{equation}
where $\eta(N)$ determines the intensity of the thermal noise, being equal to $\sqrt{2D}$, $\xi(t)$ is a white noise process and $F(N,V)$ is the free energy of the system at a given voltage. We used an overdamped Langevin code~\cite{gubbiotti2019confinement} to integrate Eq~\eqref{eq:langevin} using the Euler-Maruyama algorithm~\cite{miguel2000stochastic}. 
We can compute the first passage time distributions $\rho_{w}(\tau)$ and $\rho_{d}(\tau)$ of the wetting/drying transitions by running the simulation multiple times, initializing the system in the wet/dry state and recording the time each simulation takes to reach the other state. 
Both methods give very similar results, and we plot the comparison in Supp.~Fig.~\ref{fig:rates}.

%%%%%%%%%%%%%%%%%%%%%%%%%%%%%%%%%%%%%%%%%%%%%%%%%%%%%%%5
\newpage
\section*{Supplementary Note 3.\\
\large{Macroscopic model based on classical nucleation.}}

We used a macroscopic model to be able to estimate the parameter space for the design of an hydrophobically gated memristor. As written in the main text, we define 4 conditions that are required for an experimentally feasible HyMN:
\begin{enumerate}
\item The pore must be preferentially dry at $\Delta V = 0$;
\item The pore must be wet before a certain voltage threshold $\Delta V^*$;
\item The pore must dry out "quickly" at $\Delta V=0$ to ensure a observable transition from the wet state to the dry state;
\item The pore must wet "quickly" at the voltage threshold $\Delta V^*$
to ensure a fast transition from the dry state to the wet state.
\end{enumerate}
All the conditions can be evaluated if one as access to the free energy profile associated with the wetting variable. Computing the whole profile for all the parameter space that we explored is too computationally expensive and we have resorted to some approximations to be able to partially estimate the profile.
The first condition is met if the free energy associated with the dry state is lower than the free energy associated with the wet state. Classical nucleation theory, enhanced with the line tension, gives an expression for what is expected to be the difference of free energies between the wet and dry state, $\Delta \Omega$ \cite{giacomello2020bubble}:

\begin{equation}
    \Delta \Omega = v_p \Delta p + \gamma_{lv} A_{lv} + \gamma_{lv}\cos(\theta)A_{sv}+ \tau K,
\end{equation}
where $v_p$ is the volume of the pore filled with vapor, $\Delta P$ is the pressure difference between liquid and vapour, $\gamma_{lv}$ is the surface tension of the liquid, $A_{lv}$ is the area of the liquid-vapor interface, $\theta$ is the liquid contact angle with the solid material, $A_{sv}$ is the area of the solid-vapor interface, $\tau$ is the line tension and $K$ is the triple line length. We consider that $A_{lv}$ is twice the area of a circle,  $2\pi r_p^2$, with $r_p$ the pore radius, which is a simplification, as the area of the spherical cap of intruded liquid is slightly bigger than the area of its circular base. We consider that $A_{sv}$ is equal to $2\pi r_pl_p$, with $l_p$ the length of the pore, as this is the interior area of the pore surface. We consider K to be twice the perimeter of the pore mouth. The liquid surface tension is considered to be that of water, the pressure difference is considered to be atmospheric pressure and the contact angle is varied in this study. The value of the line tension, $\tau$, is debated in the community \cite{Tinti2023}, and in this case was a negative value of c.a -10 pN was used. 
%We report in Supp.~Fig.~\ref{fig:linetension}, the shape of the interest region for the cases of -5 pN and -15 pN.

The second condition expands on this expression by considering the effect of applied voltage. Considering the results of Supplementary Note S1, we consider that a second order term on the voltage must enter on the free energy of the dry state. This term is also proportional to the difference between the capacitance of the two states, $-\Delta C \, V^2$, with $\Delta C = C_{wet}-C_{dry}$. The capacitance of these two states is  computed as the capacitance of a planar capacitor, $\epsilon\frac{A}{l}$, where A is the area of the pore mouth, $\pi/4 d^2$, d is the diameter of the pore and l is the length of the pore and $\epsilon$ is the electric permittivity. In the dry case, we use the electric constant of vacuum, $\epsilon = \epsilon_0 = 8.852 \cdot 10^{-12}F/m^{-1}$, while in the wet case a dielectric (water) is used, and $\epsilon = \epsilon_r\epsilon_0=80 \epsilon_0$.
The third and forth conditions require estimates of the respective wetting and drying barriers. The drying barrier, $\Delta\Omega_{dry}$, is estimated using the expression used to explain experimental results of extrusion in hydrophobic nanopores \cite{lefevre2004intrusion}:
\begin{equation}
    \Delta\Omega_{dry} = \Delta P K1(\theta) r_p^3+\gamma_{lv} K2(\theta) + \tau K3(\theta) r_p,
\end{equation}
where the expressions for K1,K2,and K3 are given in detail in other work\cite{guillemot2012activated}. The wetting barrier, $\Delta\Omega_{wet}$, is obtained by summing $\Delta \Omega$ and $\Delta _{dry}$. We then use Arrhenius law to compute the respective times that take for the pores to wet, $t_{wet}$, or dry, $t_{dry}$:
\begin{equation}
    t_{wet/dry} = t_0 e^{\frac{\Delta\Omega_{wet/dry}}{k_BT}},
\end{equation}
where $t_0$ is a prefactor normally considered to be in the order of picoseconds \cite{tinti2017intrusion,lefevre2004intrusion}. We used 10 ps for our estimates.

%%%%%%%%%%%%%%%%%%%%%%%%%%%%%%%%%%%%%%%%%%%%%%%%%%%%%%%5
\newpage
\section*{Supplementary Note 4.\\
\large{Analysing the experimental voltage cycles of the engineered FraC nanopore.}}

In this section, we report the analysis done on the experimental measurements
of the current-voltage cycles reported on Fig.~3d. 
These cycles are computed by continuously measuring the current that passes through a single nanopore, 
when it is submitted to a sawtooth voltage signal, and averaging over multiple realizations after removing the capacitive current. For simplicity, we will discuss this process for the cycles done at frequency 5Hz and, but the same process was done  for the experiments reported in Figure 4.

In Supp.~Fig.~\ref{fig:fracexperiment5} and in Supp.~Fig.~\ref{fig:fracexperiment2} we report the voltage (panel a) and current (panel b) traces of a single realization of a voltage cycling experiment. The experiment consists of a sawtooth signal, and we use only the results of the 9 cycles highlighted in red. To enhance the statistics, this protocol is repeated multiple times, and the averages done correspond to at least 30 cycles. 
As can be seen in panels c of Supp.~Fig.~\ref{fig:fracexperiment5} and Supp.~Fig.~\ref{fig:fracexperiment2}, the capacitive current leads to current-voltage cycles that do not go through 0. Because we want to characterize the memristive behaviour of the FraC HyMN this capacitive current needs to be removed. 

In Supp.~Fig.~\ref{fig:fraccapacitance5} and in Supp.~Fig.~\ref{fig:fraccapacitance2} we report he respective capacitive currents (panel a), which is obtained by averaging over a voltage cycling realization (9 cycles) where the pore did not open. This allows us this current as a baseline, and if we remove it from the experimental data, the current-voltage cycles go through 0, see panel b. We are then able to average the current-voltage cycles, observing a pinched hysteresis loop (panel c).

\newpage

%%%%%%%%%%%%%%%%%%%%%%%%%%%%%%%%%%%%%%%%%%%%%%%%%
\newpage

%%%%%%%%%%%%%%%%%%%%%%%%%%%%%%%%%%%%%%%%%%%%%%%%%
\newpage

%%%%%%%%%%%%%%%%%%%%%%%%%%%%%%%%%%%%%%%%%%%%%%%%%
\newpage
\begin{figure}[h!]
    \centering
    \includegraphics[width=\linewidth]{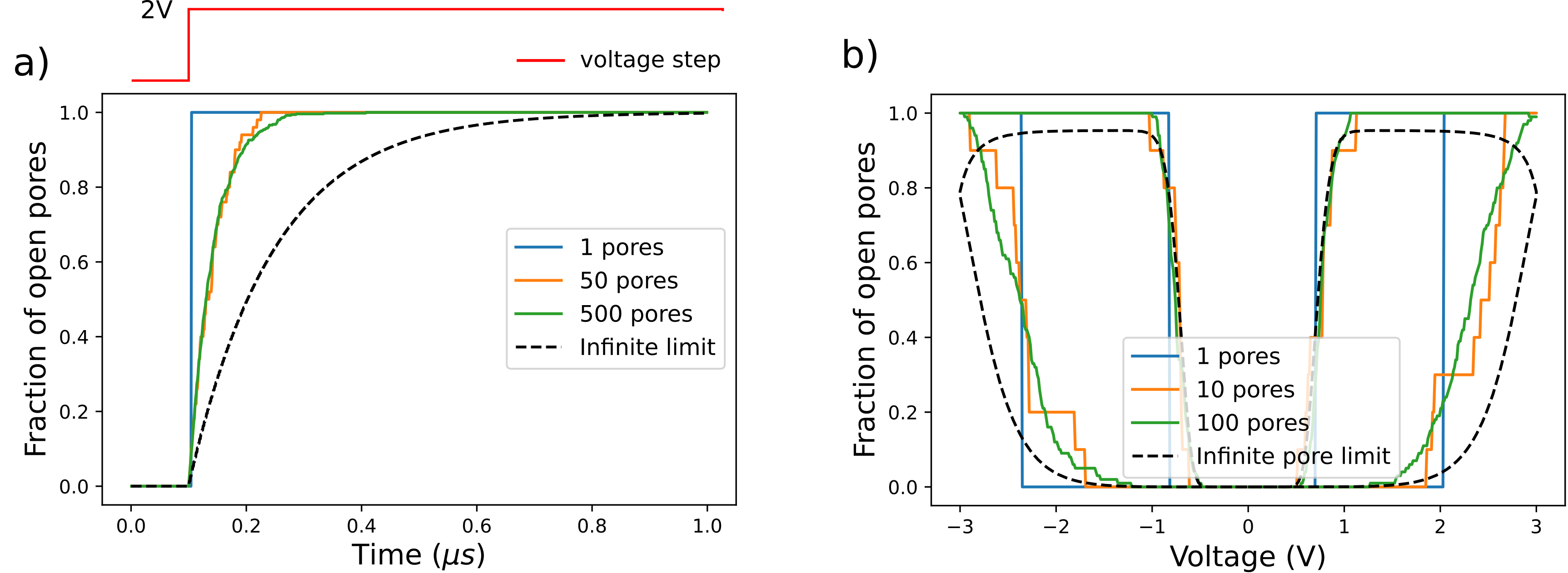}
    \caption{
    \textbf{Collective pore wetting.}
    a) Response of different number of pores, 1, 50 and 500,
    when a step voltage is applied at $t_0 = 0.1$~$\mu$s, see full black line.
    Black dashed line is computed integrating Eq~2 of the main text, 
    and represents the behaviour of an infinite number of pores.
    Colored lines are computed via a Langevin Dynamics
    see Supp.~Fig.~\ref{fig:rates}.
    b) Voltage cycles, computed via the same protocol as the one in
    Fig.~1 of the main text, for different number of pores, 1, 10 and 100.
    The stochastic opening of the pores is smoothed out as the number of pore increases, and just some tens of pores, between 10 and 50, are enough for a smooth behaviour.
    It must be noted that the limit behavior for the 
    Langevin simulations and the infinite limit computed via Eq~2 is not the same, 
    due to the fact that we are using the rates computed with Eq~\ref{eq:ratetheory}, 
    and they are slightly different from the ones computed with Langevin.
    }
    \label{fig:collective-wet}
\end{figure}

%%%%%%%%%%%%%%%%%%%%%%%%%%%%%%%%%%%%%%%%%%%%%%%%%
\newpage
\begin{figure}[h!]
    \centering
    \includegraphics[width=\linewidth]{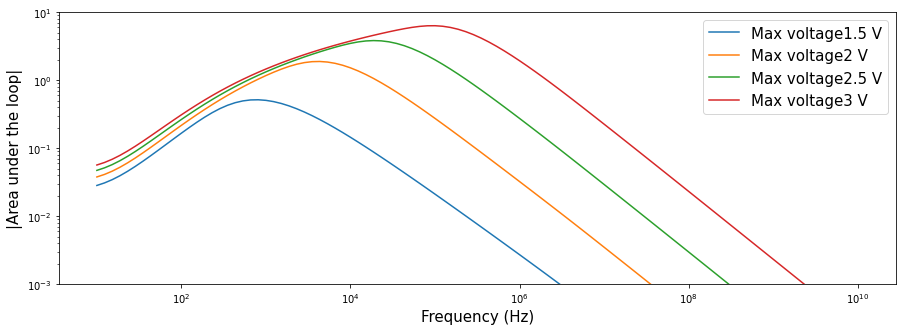}
    \caption{
    \textbf{Dependence of the area under the hysteresis curve with the maximum applied voltage. }
    The hysteresis loop observed in Fig.~1 of the main text, 
    and the analysis of the dependence of the area under the loop on the frequency, were done using a given maximum applied voltage (3V, corresponding to 2.5 $V_c$). For any given maximum applied voltage there exists a frequency that maximizes the are under the curve. In the above panel, we show how the maximizing frequency increases as the maximum applied voltage increases. The area under the curve also increases with the applied voltage. Of course even though in theory these peaks can move while increasing or decreasing the applied voltage there is a minimum current that can be measured, bounding the frequencies from below, and a maximum cycling speed that can be done, bounding the frequencies from above. Low maximum applied voltage, below $V_c$, would create almost indistinguishable hysteresis, and high maximum applied voltage would require very high cycling frequencies for hysteresis to be observed.
    }
    \label{fig:areacurve}
\end{figure}

%%%%%%%%%%%%%%%%%%%%%%%%%%%%%%%%%%%%%%%%%%%%%%%%%
\newpage
\begin{figure}[h!]
    \centering
    \includegraphics[width=\linewidth]{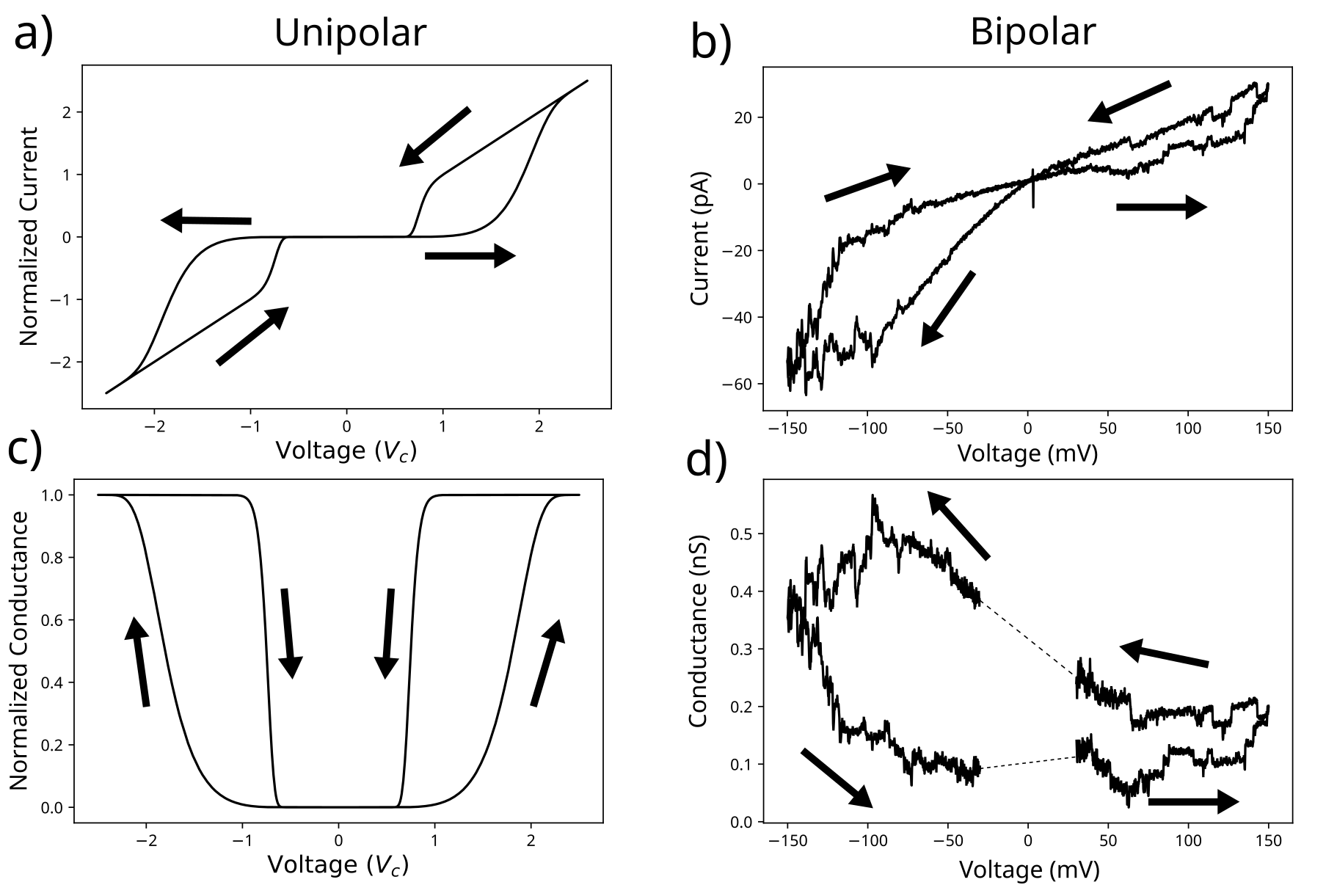}
    \caption{
    \textbf{Type of memristor behaviour.}
    In this figure, we discuss the different type of memristor behaviour observed on hydrophobically gated memristive nanopores (HyMNs). In a) we plot the I-V curve represented in Fig.1d) of the main text, indicating the direction of the voltage cycle. The current does not self intersect at the origin, but the conductance of the system does, see panel c). This is the characteristic behaviour of an unipolar memristor. In panel b) we plot the experimental I-V curve represented in Fig.3d) of the main text, indicating the direction of the voltage cycle. The current self intersects at the origin, but the conductance of the system does not, see panel d). This is the characteristic behaviour of a bipolar memristor. 
    }\label{fig:memtristortype}
\end{figure}

%%%%%%%%%%%%%%%%%%%%%%%%%%%%%%%%%%%%%%%%%%%%%%%%%
%\newpage
%\begin{figure}[h!]
%    \centering
%    \includegraphics[width=\linewidth]{supplementary_figures/S6.png}
%    \caption{
%    Effect of the line tension. Panel (a) shows how the 4 design parameters intersect for a contact angle of 104° and a maximum wetting voltage of 0.2 V and line tension of -5 pN and (b) corresponds to a value for the line tension of -15 pN. Like in the main text, the green regions correspond to the pore dimensions, which require a voltage higher than 0.2 V to wet. The red regions correspond to the pore dimensions where wet state is more favorable than the dry state at 0V. The blue regions correspond to the pore dimensions that have the a average pore drying time of more that 10 seconds when no voltage is applied. The black regions correspond to the pore dimensions where the pore takes longer than 10 seconds to wet at 0.2V.
%    }
%    \label{fig:linetension}
%\end{figure}

%%%%%%%%%%%%%%%%%%%%%%%%%%%%%%%%%%%%%%%%%%%%%%%%%
\newpage
\begin{figure}[h!]
    \centering
    \includegraphics[width=\linewidth]{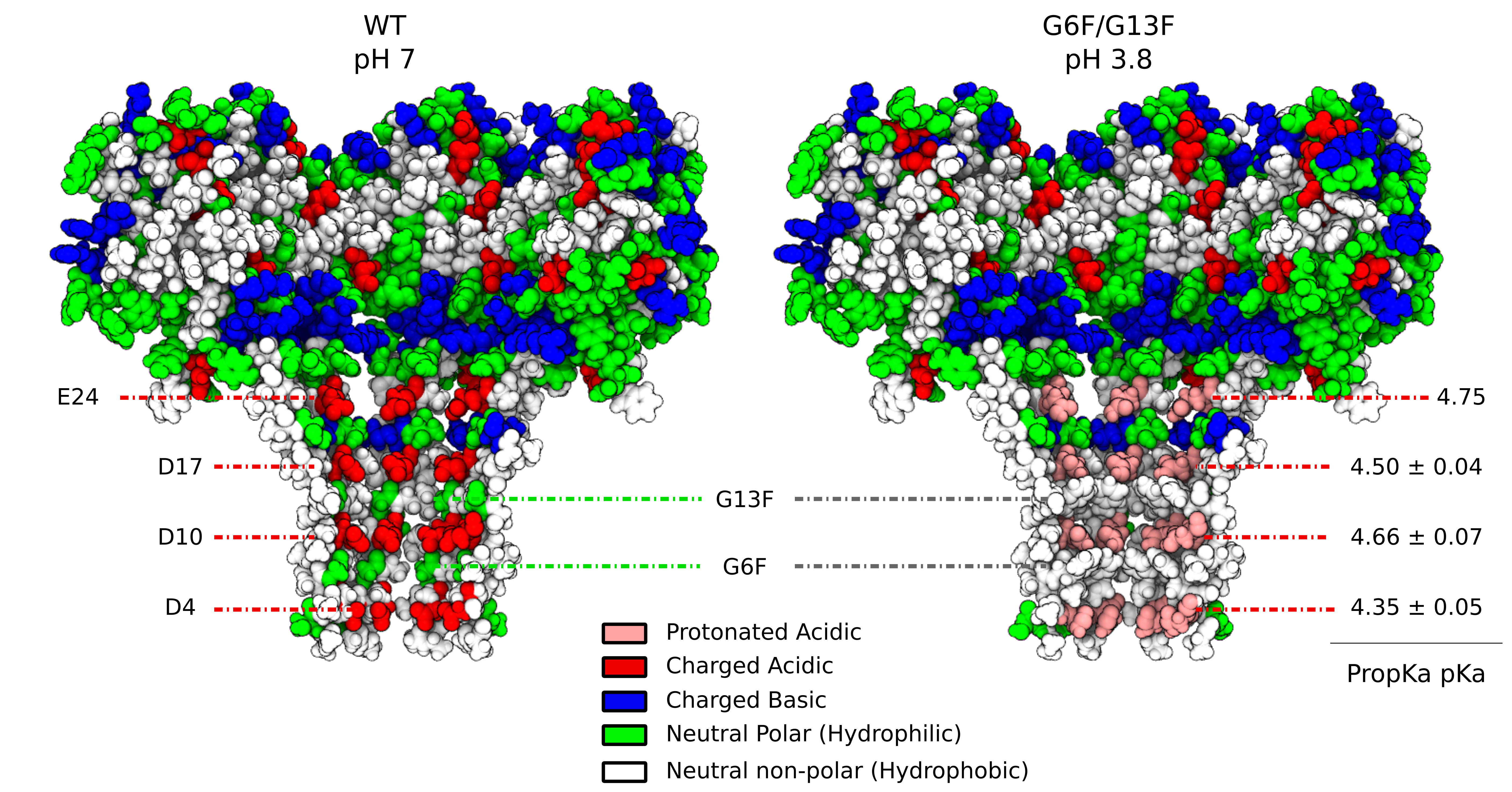}
    \caption{
    \textbf{Molecular model} of WT (left) and hydrophobically engineered (right) FraC nanopore.
    In our experiments, the glycines G6 and G13 are mutated into high hydrophobic phenilalanines (contact angle > 110°)~\cite{zhu2016characterizing}.
    At lower pH, all the acidic residues in the constriction are completely neutralized;
    in this condition, the extreme confinement together with the hydrophobicity of phenilalanine residues 
    allows the nucleation of a stable vapor bubble inside the pore constriction, see Fig.~3 of the main manuscript.
    }
    \label{fig:fracexperiment}
\end{figure}

%%%%%%%%%%%%%%%%%%%%%%%%%%%%%%%%%%%%%%%%%%%%%%%%%
\newpage
\begin{figure}[h!]
    \centering
    \includegraphics[width=0.70\linewidth]{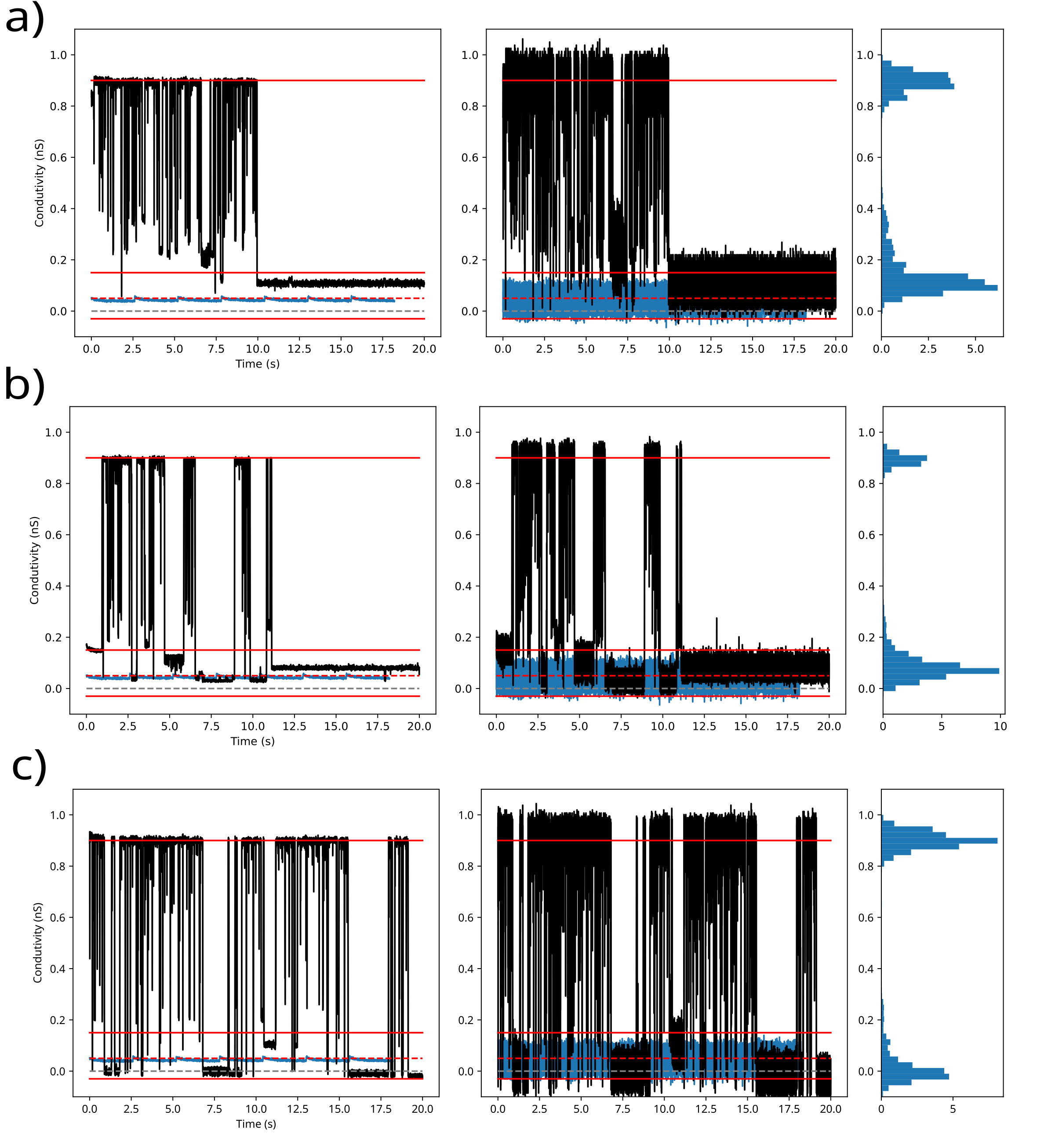}
    \caption{\textbf{Single pore current traces at constant applied voltage and pH 3.8.}
    \tr{Pannels a, b and c correspond to traces at constant applied voltage, -25 mV, -50 mV and 50 mV respectivelly. In these pannels, the first graph corresponds to the recorded signal convoluted with a constant signal, in such a way as to do a moving average with a window of 0.01s. The second graph correspond to the raw signal measured. Red lines were added at 0.9 nS, the open pore conductance, and 0.15 nS the initial state of the pore at -50 mV, and at -0.05 ns the bottom conductance state at 50 mV and are the same in all the figures. The blue signal corresponds to the signal of a membrane with no pores, and the red dashed line corresponds to its average value. The grey dashed line corresponds to 0. For each of the raw signal measurement we have included an histogram for the measured conductance, which shows two clear peaks for a low conductance and an high conductance state. The data present in panel c is the same as in figure 3 c of the main text.}}
    \label{fig:constantvoltage}
\end{figure}

%%%%%%%%%%%%%%%%%%%%%%%%%%%%%%%%%%%%%%%%%%%%%%%%%
\newpage
\begin{figure}[h!]
    \centering
    \includegraphics[width=\linewidth]{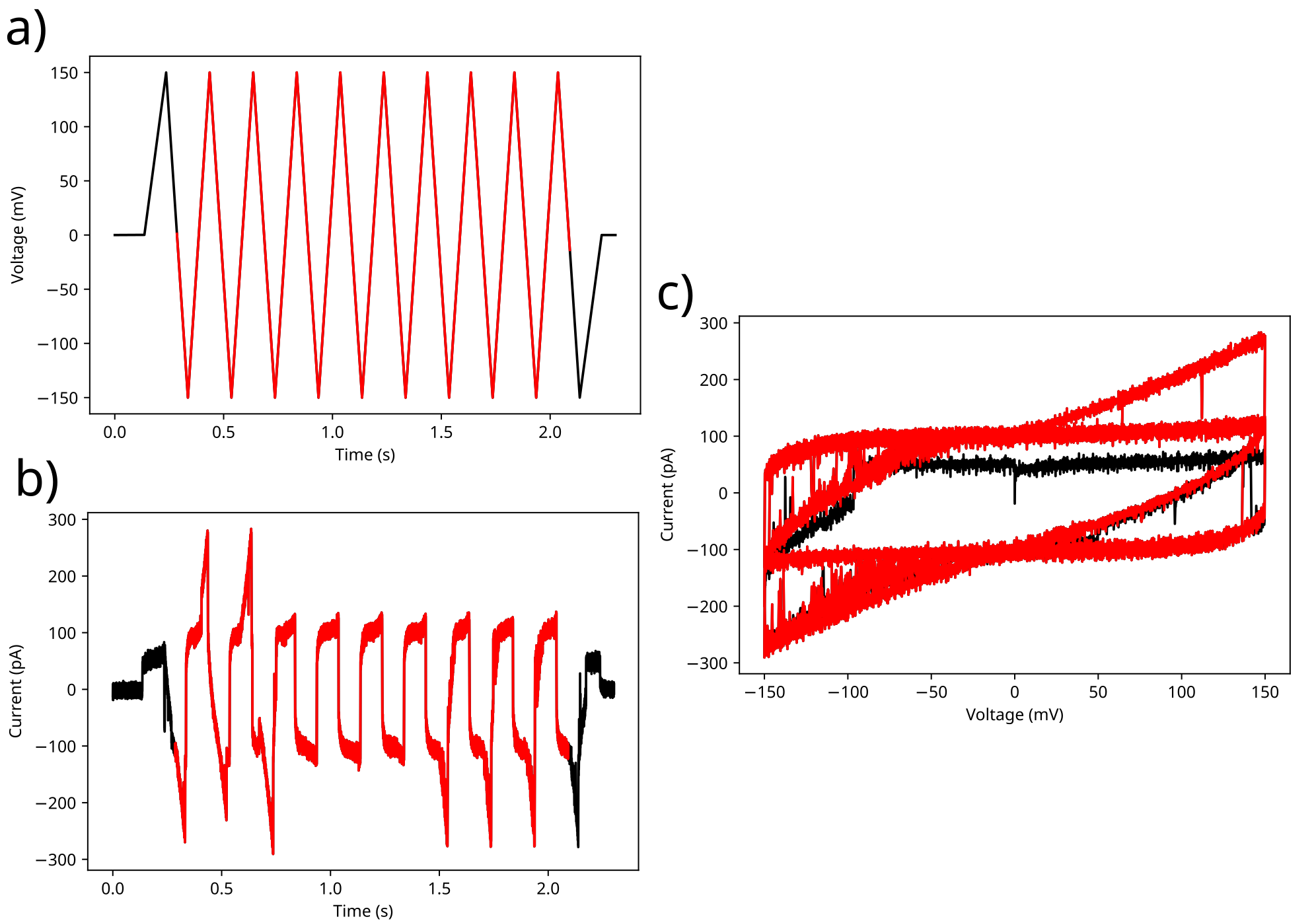}
    \caption{
    \textbf{Experimental current signal of a voltage cycle of 5Hz through FraC nanopore at pH 3.8. }
    Panel a) corresponds to the imposed voltage trace at 5Hz, where the voltage is changed from 150 mV to -150mV and back. Panel b) corresponds to the current measured during a single realization of this experiment. The characteristic signal of the capacitive current can be observed, while the signal corresponding to the opening of the pore gives rise to "spikes" on top of that current. Panel c) shows the current-voltage curve, which does not cross zero, with a non-pinched hysteresis curved due to the capacitive current. In all panels, in red are highlighted the 9 cycles used for all the averages. In this figure, we show a single realization of the data used in figure 3d.}
    \label{fig:fracexperiment5}
\end{figure}
%%%%%%%%%%%%%%%%%%%%%%%%%%%%%%%%%%%%%%%%%%%%%%%%%
\newpage
\begin{figure}[h!]
    \centering
    \includegraphics[width=\linewidth]{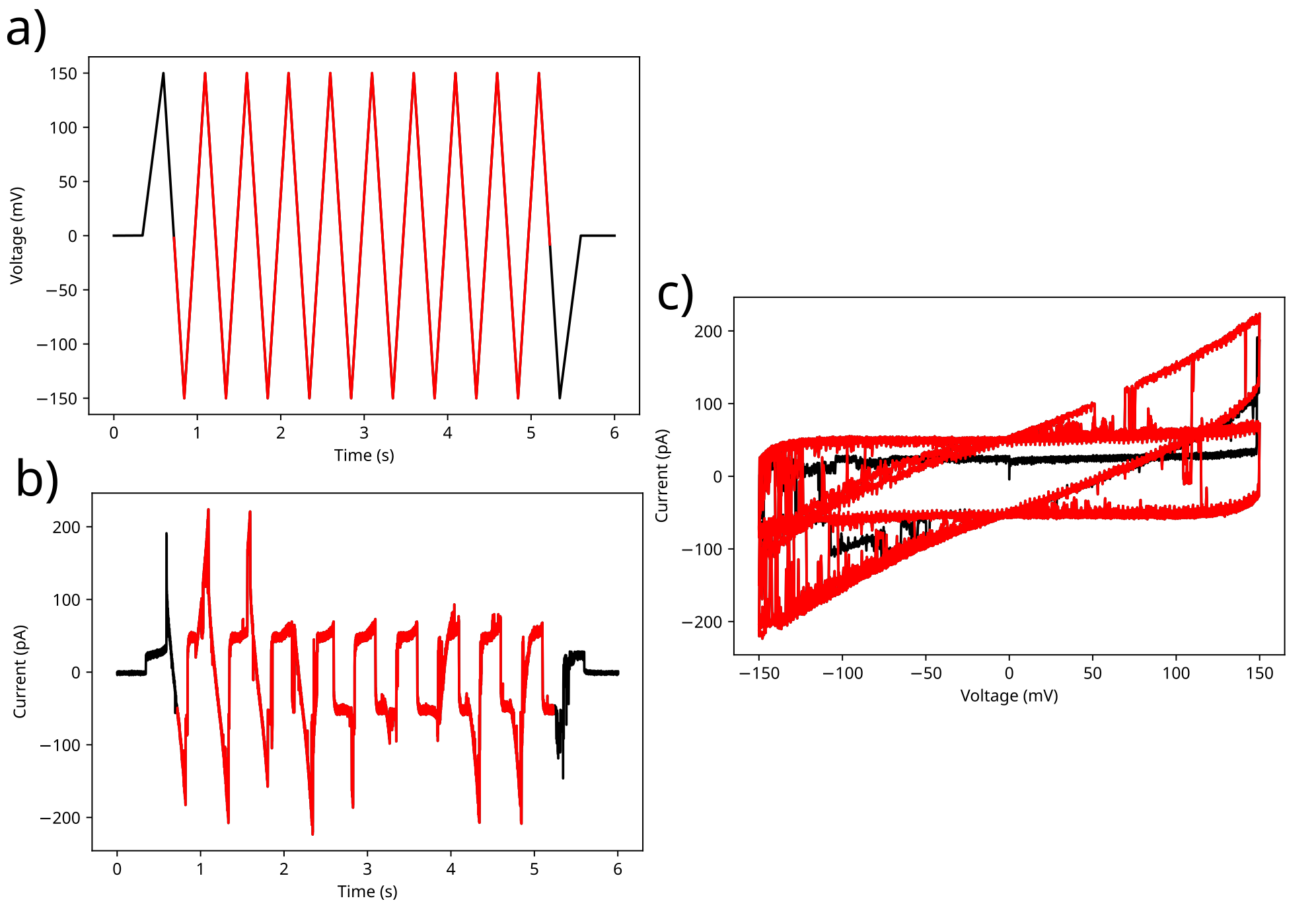}
    \caption{
    \textbf{Experimental current signal of a voltage cycle of 2Hz through FraC nanopore at pH 3.8. }
    Panel a) corresponds to the imposed voltage trace at 2Hz, where the voltage is changed from 150 mV to -150mV and back. Panel b) corresponds to the current measured during a single realization of this experiment. The characteristic signal of the capacitive current can be observed, while the signal corresponding to the opening of the pore gives rise to "spikes" on top of that current. Panel c) shows the current-voltage curve, which does not cross zero, with a non-pinched hysteresis curved due to the capacitive current. In all panels, in red are highlighted the 9 cycles used for all the averages. In this figure, we show a single realization of the data used in figure 3d.}
    \label{fig:fracexperiment2}
\end{figure}

%%%%%%%%%%%%%%%%%%%%%%%%%%%%%%%%%%%%%%%%%%%%%%%%%
\newpage
\begin{figure}[h!]
    \centering
    \includegraphics[width=\linewidth]{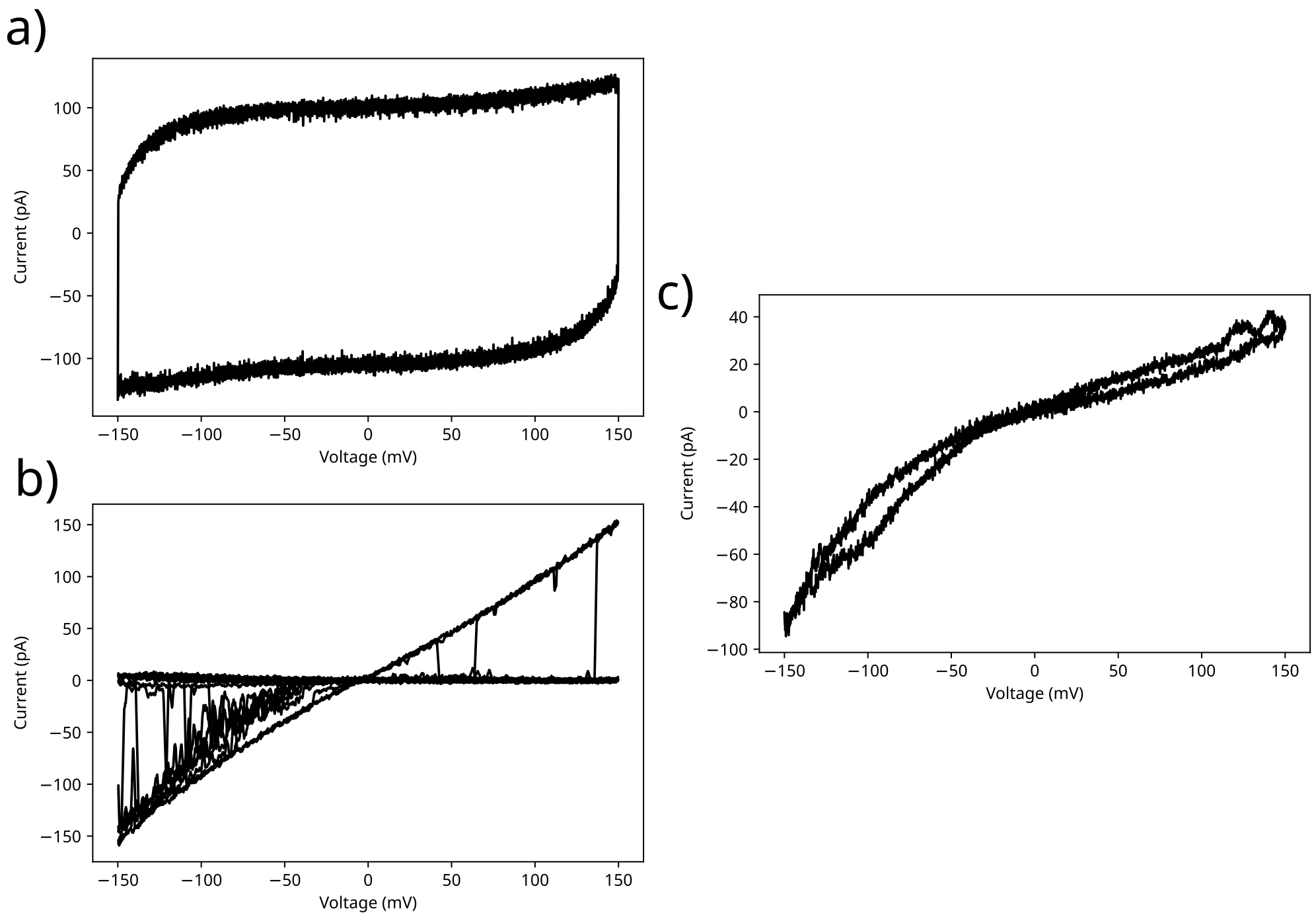}
    \caption{\textbf{Removing the capacitive current at 5Hz and pH 3.8.} Panel a) corresponds to the average current over 9 voltage cycles where the pore never opened. The measured current is used as the baseline to remove from the other realizations. Panel b) corresponds to the subtraction of the measured current, shown in red in figure S9, and the capacitive current, with the current signal now going through 0. Panel c) shows the average current-voltage curve of a single realization, by averaging out ca. 100 voltage cycles where the pore opened at least one. This figure corresponds to the same data as figure S9 and figure 3d of the main text.  }
    \label{fig:fraccapacitance5}
\end{figure}

%%%%%%%%%%%%%%%%%%%%%%%%%%%%%%%%%%%%%%%%%%%%%%%%%
\newpage
\begin{figure}[h!]
    \centering
    \includegraphics[width=\linewidth]{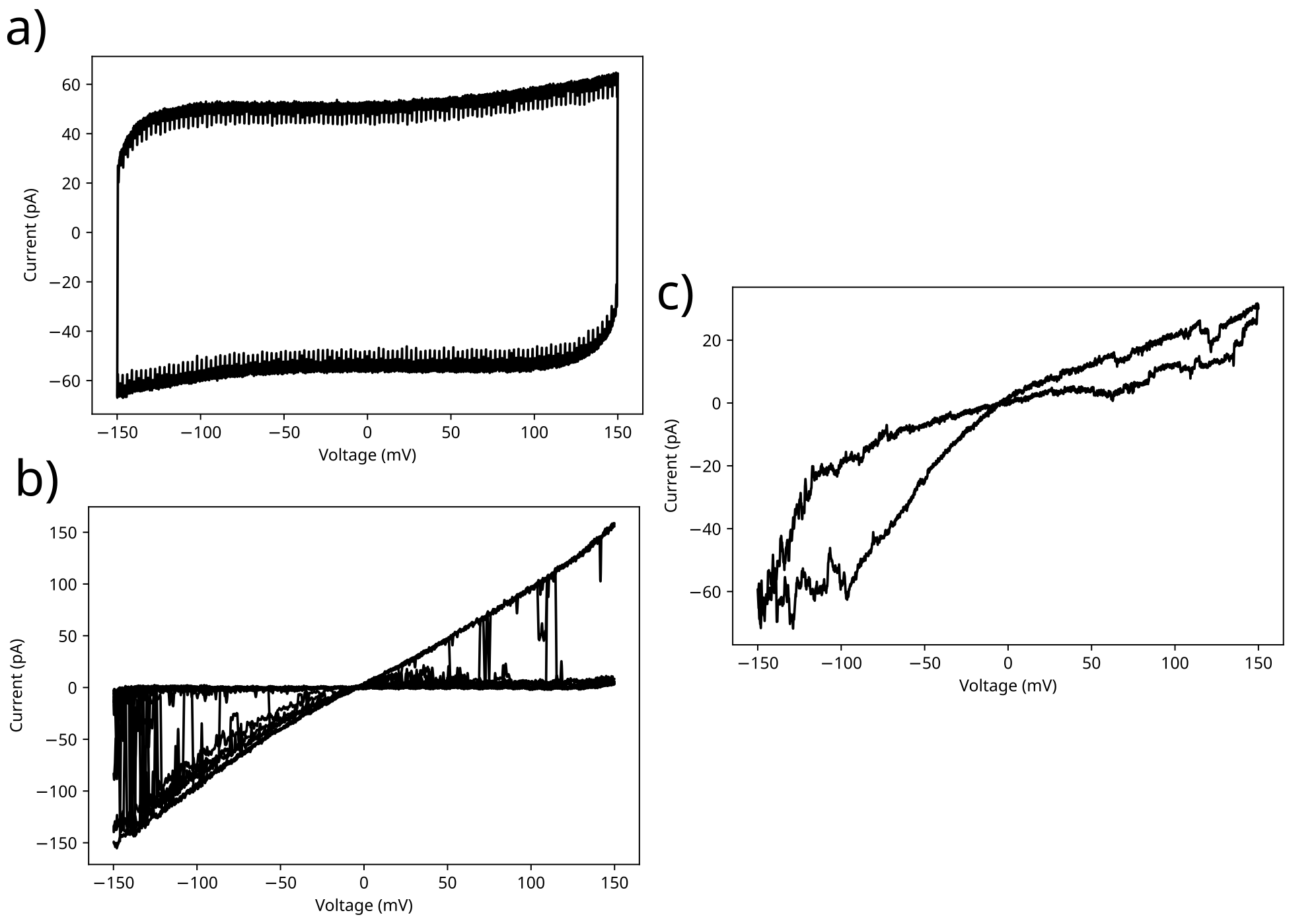}
    \caption{\textbf{Removing the capacitive current at 2Hz and pH 3.8.} Panel a) corresponds to the average current over 9 voltage cycles where the pore never opened. The measured current is used as the baseline to remove from the other realizations. Panel b) corresponds to the subtraction of the measured current, shown in red in figure S10, and the capacitive current, with the current signal now going through 0. Panel c) shows the average current-voltage curve of a single realization, by averaging out ca. 40 voltage cycles where the pore opened at least one. This figure corresponds to the same data as figure S10 and figure 3d of the main text.  }
    \label{fig:fraccapacitance2}
\end{figure}

%%%%%%%%%%%%%%%%%%%%%%%%%%%%%%%%%%%%%%%%%%%%%%%%%
\newpage
\begin{figure}[h!]
    \centering
    \includegraphics[width=\linewidth]{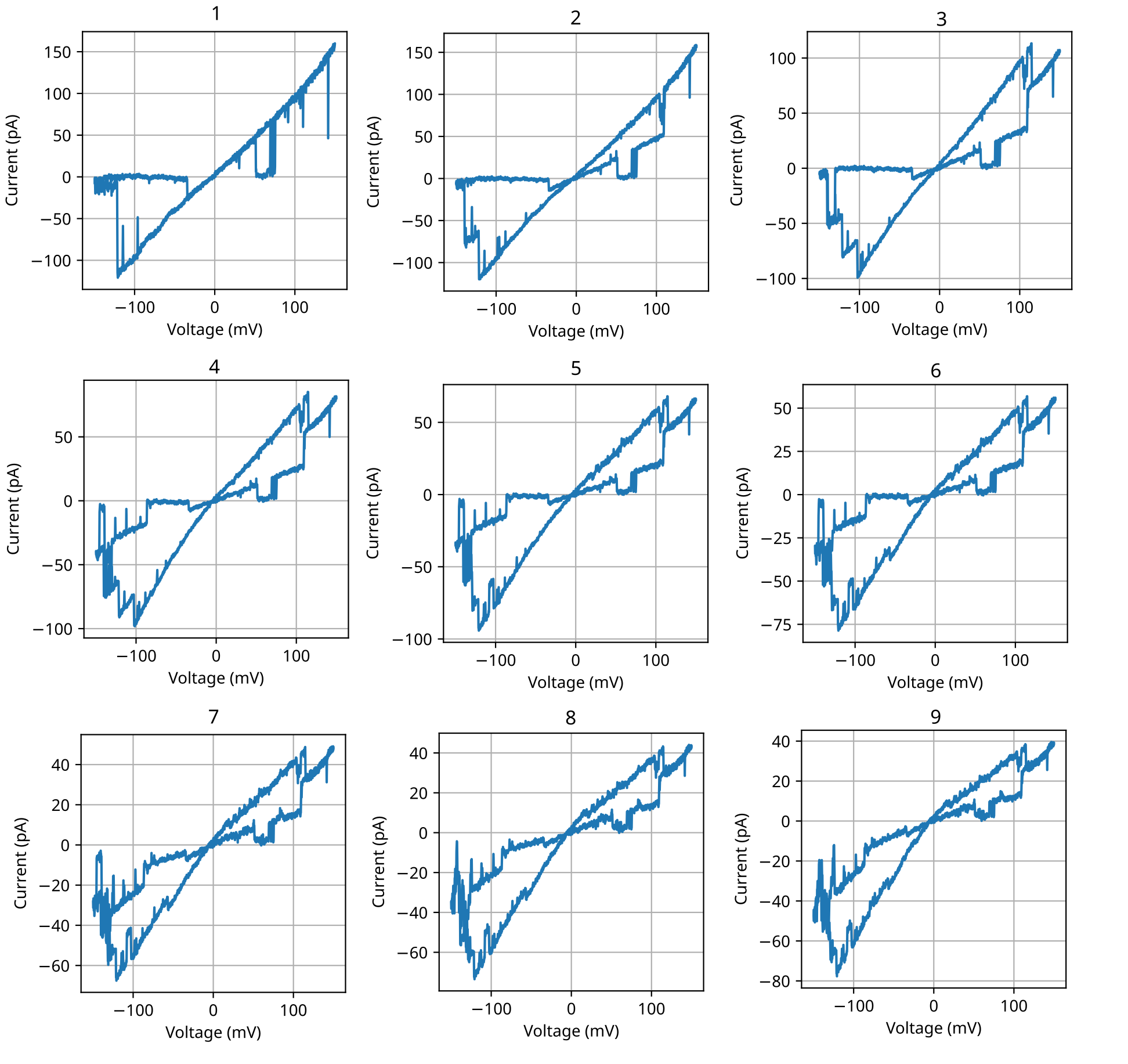}
    \caption{\textbf{From single channel trace to average memristive behaviour at 2Hz and pH 3.8.}
    \tr{Panel 1 corresponds to a single voltage cycle measurement of a single pore at 2Hz and at pH 3.8. Panels 2 to 9 correspond to averaging the signal for that number of consecutive cycles, e.g. panel 9 corresponds to the averaging of 9 consecutive cycles. Single channel recordings have the trace of stochastic memristors, and averaging the cycles makes the traditional pinched loop stand out. This data corresponds to the same data shown in figures S10, S12 and figure 3d of the main text. }
    }
    \label{fig:fracaveraging}
\end{figure}

%%%%%%%%%%%%%%%%%%%%%%%%%%%%%%%%%%%%%%%%%%%%%%%%%
\newpage
\begin{figure}[h!]
    \centering
    \includegraphics[width=\linewidth]{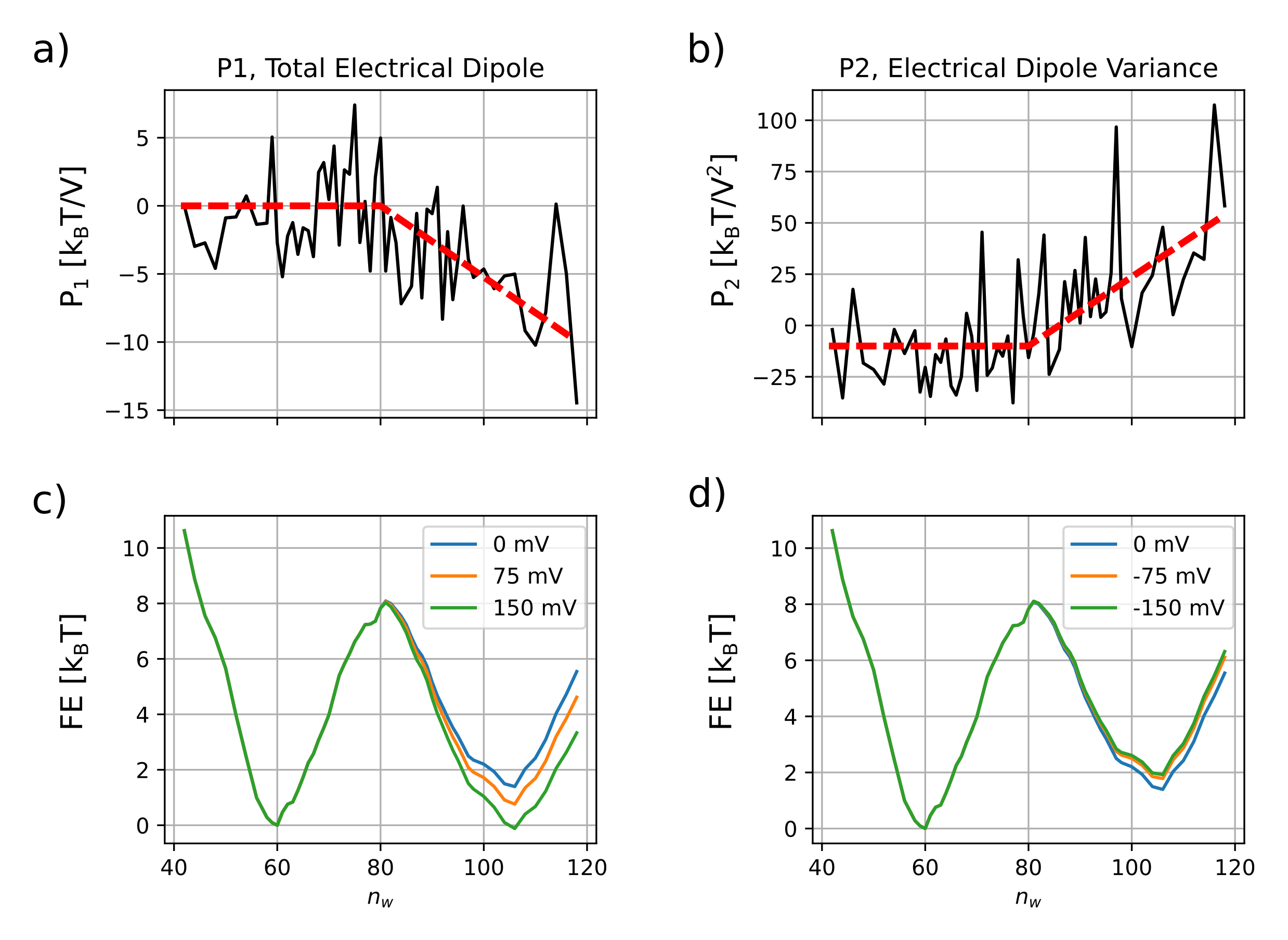}
    \caption{
    \textbf{Effect of the voltage on the free energy profile of FraC nanopore.}
    Values of the first order coefficient, P1 (panel a), and the second order coefficient, P2 (panel b) on the expansion of the free energy discussed in Supp.~Note~S1, Eq.S25, obtained from the RMD simulations of the FraC system 
    shown in Fig.~3 of the main manuscript. The P1 and P2 value used in the expansion are the averaged profiles shown by the red dashed lines.
    Panel c) show the free energy profile changes at positive applied voltages,
    while panel d) shows it at negative voltages.
    It can be noted the asymmetric tilting of the free energy profile under opposite voltages:
    at positive voltages,
    the model predict that the wet and dry states becomes equiprobable at around
    150~mV, while keeping the wetting barrier the same.
    Differently, at negative voltages the dry state remains the most probable up to -150~mV;
    \tr{
    moreover, it is interesting to note that our theory predicts, at negative voltages, that the electric field reduces the drying barrier, 
    in opposition to what is commonly expected in electrowetting.
    This is due to the interplay of intrinsic dipoles present into the system and the external electric field. However, a comprensive analysis of this "electrodrying" phenomena will require a dedicated study.
    }
    }
    \label{fig:fracfreeenergy}
\end{figure}

\newpage

\bibliography{biblio.bib}